\documentclass[sigconf, screen]{acmart}

\usepackage{subfig,graphicx}
\usepackage{listings}
\usepackage{amsmath}
\usepackage[noend]{algpseudocode}
\usepackage{physics}
\usepackage{multirow}
\usepackage{enumitem}
\newcommand{\ignore}[1]{}
\usepackage{wrapfig}
\usepackage{braket}
\usepackage{mathtools}
\usepackage{balance}
\usepackage{tabularx}

\usepackage{url}

\usepackage{float}

\usepackage{xparse}

\usepackage{graphicx}

\usepackage{empheq}
\usepackage{float}

\usepackage{xspace}

\usepackage{subfig}
\NewDocumentCommand{\tensor}{t_}
 {%
  \IfBooleanTF{#1}
   {\tensop}
   {\otimes}%
 }
\usepackage{makecell}
\usepackage[section]{placeins}

\usepackage{textcomp}

\def \qtree {TQSim\xspace}

\def \minSpeedup {1.59$\times$\xspace}
\def \maxSpeedup {3.89$\times$\xspace}
\def \averageSpeedup {2.51$\times$\xspace}

\def \averageFidelityDiff{0.006\xspace}
\def \maxFidelityDiff {0.016\xspace}

\usepackage{cleveref}

\usepackage{xcolor}

\renewcommand\fbox{\fcolorbox{blue}{white}}

\newcommand{\finding}[1]{
\vspace{0.5ex}
\noindent\fbox{\begin{minipage}{0.975\linewidth}
{\sf\bf\it #1}
\end{minipage}}
}

\ccsdesc[500]{Hardware~Quantum computation}
\ccsdesc[500]{Computing methodologies~Quantum mechanic simulation}

\keywords{Quantum Computing, Noisy Monte Carlo Simulation, State Reuse}

\copyrightyear{2025}
\acmYear{2025}
\setcopyright{acmlicensed}\acmConference[ISCA '25]{Proceedings of the 52nd Annual International Symposium on Computer Architecture}{June 21--25, 2025}{Tokyo, Japan}
\acmBooktitle{Proceedings of the 52nd Annual International Symposium on Computer Architecture (ISCA '25), June 21--25, 2025, Tokyo, Japan}
\acmDOI{10.1145/3695053.3730992}
\acmISBN{979-8-4007-1261-6/2025/06}

\begin{document}

\title{Accelerating Simulation of Quantum Circuits under Noise via Computational Reuse}

 \author{Meng Wang}
 \orcid{0009-0008-1749-7929}
 \affiliation{%
   \institution{The University of British Columbia}
   \city{Vancouver}
   \country{Canada}}
 \email{mengwang@ece.ubc.ca}

 \author{Swamit Tannu}
 \orcid{0000-0003-4479-7413}
 \affiliation{%
   \institution{University of Wisconsin--Madison}
   \city{Madison}
   \country{USA}}
 \email{swamit@cs.wisc.edu}

 \author{Prashant J. Nair}
 \orcid{0000-0002-1732-4314}
 \affiliation{%
   \institution{The University of British Columbia}
   \city{Vancouver}
   \country{Canada}}
 \email{prashantnair@ece.ubc.ca}

\renewcommand{\shortauthors}{Wang et al.}

\begin{abstract}
To realize the full potential of quantum computers, we must mitigate qubit errors by developing noise-aware algorithms, compilers, and architectures. Thus, simulating quantum programs on high-performance computing (HPC) systems with different noise models is a de facto tool researchers use. Unfortunately, noisy simulators iteratively execute a similar circuit for thousands of trials, thereby incurring significant performance overheads.

To address this, we propose a noisy simulation technique called Tree-Based Quantum Circuit Simulation (\qtree)~\footnote{Publicly available at \url{https://github.com/meng-ubc/TQSim}.\\Archived version at \url{https://doi.org/10.5281/zenodo.15104095}.}. \qtree exploits the reusability of intermediate results during the noisy simulation, reducing computation. \qtree dynamically partitions a circuit into several subcircuits. It then reuses the intermediate results from these subcircuits during computation. Compared to a noisy Qulacs-based baseline simulator, \qtree achieves a speedup of up to \maxSpeedup for noisy simulations. \qtree is designed to be efficient with multi-node setups while also maintaining tight fidelity bounds.
\end{abstract}

\maketitle

\section{Introduction}
\label{Introduction}

Quantum computers can solve problems in minutes that would take supercomputers days~\cite{supremacy,advantage,zuchongzhi,morello2018would,quest}. However, access to real machines is limited and costly, with only a few publicly available~\cite{IBM-QCS,kumar2025contextswitchingsecuremultiprogramming,qoncord,micro3}. Quantum circuit simulation helps speed up and validate quantum computing research~\cite{Quantum_speedup_estimation}. Yet, simulating gate-based circuits presents a hurdle due to exponential state space expansion with qubit count using state vector-based simulation. Incorporating noise adds further complexity, requiring the emulation of multiple noise-injected trials or `shots.' Our evaluations (Table~\ref{table:simulation-time}) show that even simulating a 20-qubit circuit on a 32-core CPU can take thousands of seconds (despite using minimal memory).

\begin{figure}[t]
    \centering
    \includegraphics[width=\columnwidth]{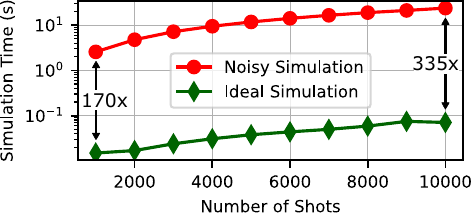}
   \caption{Simulation times (in seconds) for ideal and noisy 15-qubit Quantum Fourier Transform (\texttt{QFT}) circuits, using two 16-core Intel\textregistered\ Xeon\textregistered\ 6130 processors. Noisy simulations are 170$\times$ to 335$\times$ slower than ideal ones.}
    \label{fig:base_ex_time}
\end{figure}
 
\begin{figure*}%
    \centering
    
    \subfloat[\centering (a) Ideal simulation and noisy operators.]
    {
    {\includegraphics[width=0.25\linewidth]{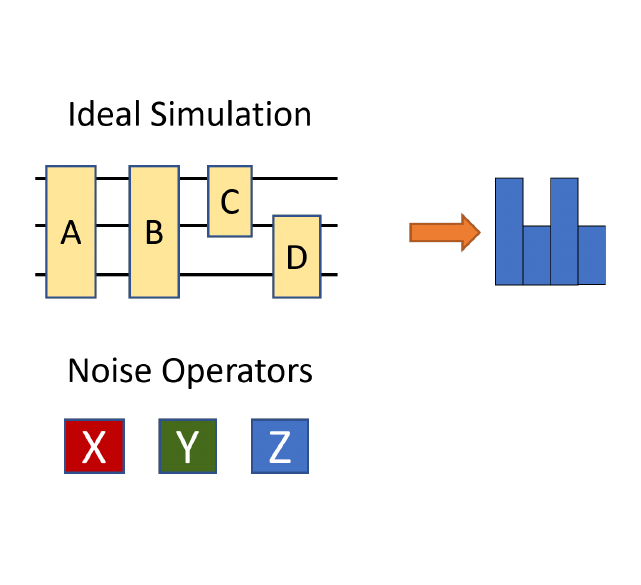} }
     \label{fig:overview_ideal}}%
     \hspace{0.05in}
    \subfloat[\centering (b) Flow of the baseline state vector simulator.]
    {{\includegraphics[width=0.3\linewidth]{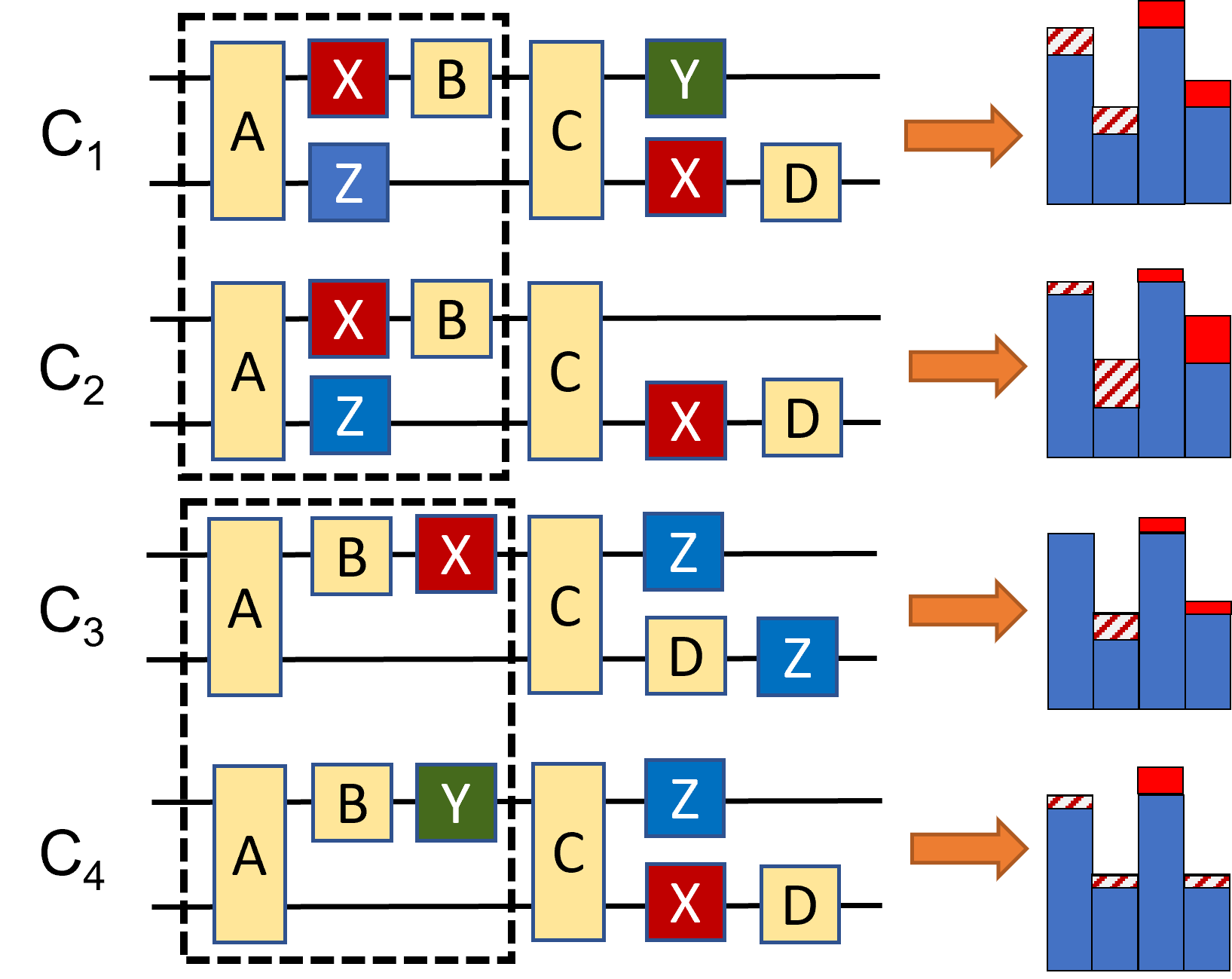} }
     \label{fig:overview_noisy_baseline}}%
    \subfloat[\centering (c) Flow of \qtree.]
    {{\includegraphics[width=0.35\linewidth]{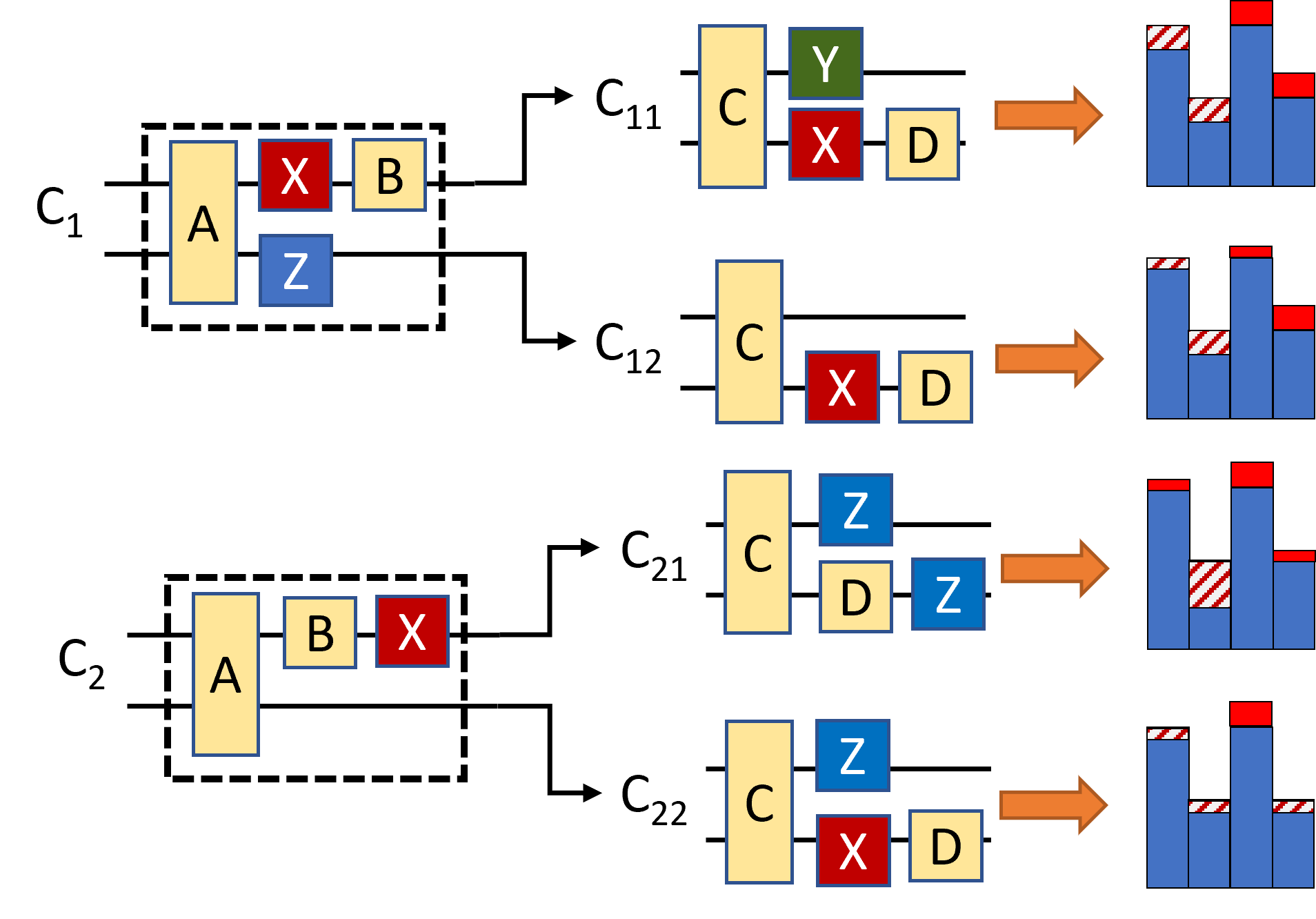} }
     \label{fig:overview_noisy_reuse}}%
    \caption{Noisy circuits and potential reuse of the intermediate states. (a) Ideal simulation and possible noise operators. (b) Four noisy circuits are generated from the original circuit, and their noisy-version resulting distributions. (c) Reuse the intermediate state after gate B and the new noisy-version resulting distributions. Note that the noise operator \texttt{Y} in $C_4$ from (b) has been replaced by a noise operator \texttt{X} in (c). This illustrates the source of loss in accuracy.}%
    \label{fig:overview_design}%
\end{figure*}
\noindent \textbf{Why Optimize Schrödinger-style Simulation?} Quantum circuits are typically simulated using either Schrödinger-style methods~\cite{schrodinger1926undulatory,schrodinger1926quantisierung} (state vector, density matrix) or Feynman path integrals~\cite{feynman1948space,feynman2010quantum}. While the Feynman method is more memory-efficient, it requires exponentially more time. Despite this, Schrödinger-style simulation is the predominant method in most advanced quantum simulation packages~\cite{qiskit2024, cirq_developers_2020_4064322} due to its flexibility and versatility in supporting different noise channels. This reflects a key priority in quantum simulation: \emph{runtime and flexibility often outweigh memory use}. By optimizing Schrödinger-style simulations, we align with this industry trend\footnote{\url{https://ionq.com/resources/the-value-of-classical-quantum-simulators}}, that is, we focus on reducing computational time - a key factor in the feasibility of quantum simulations.

Previous works mainly targeted single-shot quantum simulations~\cite{cutqc, HyQuas, hpca_faster_shrodiner_style, bsd}.  In contrast, multi-shot simulations, running circuits repeatedly, offer more room for performance gains. This paper proposes a fast and flexible simulator to improve multi-shot workloads and support exploration of NISQ-era algorithms.

\noindent \textbf{Overheads of Noisy Simulations:} Noisy quantum simulations add local unitary noise operators, acting on one or two qubits, to mimic noise effects. This causes significant slowdowns due to linear growth in computation per gate and disrupts optimizations like gate fusion, vectorization, and compute intensity~\cite{noisy_qsim_cirq,wang2023enabling,li2024tanqsimtensorcoreacceleratednoisy}.

Figure~\ref{fig:base_ex_time} shows the simulation time for the 15-qubit Quantum Fourier Transform (\texttt{QFT}) circuit using the Qulacs~\cite{qulacs} simulator. It shows ideal and noisy simulations with the depolarization noise model. The noise model uses 0.1\% and 1.5\% error rates for single and two-qubit gates, respectively~\footnote{Error rates are derived from Google Sycamore characterization~\cite{quantum_datasheet,supremacy}.}. Notably, the noisy simulation time is 170$\times$ to 335$\times$ higher than that for the ideal circuits.

\noindent \textbf{High-Level Insight and Computation Efficiency:} Figure~\ref{fig:overview_ideal} shows the noise-free circuit alongside noise operators. Figure~\ref{fig:overview_noisy_baseline} shows noisy circuits generated for a 4-shot simulation task, with each noisy circuit containing randomly inserted noise operators~\footnote{The listed noise operators are for depolarizing error channels. We evaluate \qtree with other error channels listed in Section~\ref{Methodology}, and the results are shown in Section~\ref{Evaluation}.}. This study identifies that there is potential \emph{computation reuse} within these circuits. Specifically, the circuit segments that are highlighted by dashed lines in Figure~\ref{fig:overview_noisy_baseline} share similar noisy operators. Existing simulators redundantly execute \emph{similar computations} across shots in these sections. Intermediate states can be reused to mitigate this redundancy. Figure~\ref{fig:overview_noisy_reuse} showcases the reuse of the intermediate state after gate B, which leads to a 25\% computation reduction.

\noindent \textbf{Our Proposal:} Using this insight, we introduce \qtree, a tree-based quantum circuit simulator that uses state vectors. \qtree leverages intermediate state reuse to expedite noisy quantum simulations. \qtree intelligently divides the quantum circuit into subcircuits. Each subcircuit produces an intermediate state. These intermediate states are then reused across multiple shots. For instance, the circuit in Figure~\ref{fig:overview_design} consists of two subcircuits: one with gates A and B, and another with gates C and D, where the result from the first subcircuit is reused twice. \qtree tackles two key challenges: first, efficiently partitioning the circuit and determining the optimal reuse count for subcircuits; second, generating more subcircuits can increase performance but may require more memory for storing states while also affecting outcome fidelity. \qtree uses three novel approaches that address these challenges and balance reuse and fidelity bounds.

\noindent \textbf{1. Circuit Partitioning for Reuse:} \qtree proposes a dynamic circuit partition (DCP) technique that determines the circuit partitions and shot distributions based on the noise model's error rates and the circuit's gate count. This enables DCP to obtain a significant speedup while producing results similar to those from the baseline noisy simulator. This approach is particularly useful for high-performance computing (HPC) systems, which are often used to simulate large-scale quantum circuits.

\noindent \textbf{2. Optimizing Memory Efficiencies:} Large-scale systems like Frontier, Summit, and Perlmutter predominantly rely on GPU compute nodes, leading to significant underutilization of CPU and storage memory~\cite{frontier2022,perlmutter2022,summit2018}. Our experiments demonstrated low system memory utilization: only 16.7\% on Frontier, 5.3\% on Summit, and 30.8\% on Perlmutter. This underutilization arises from two main factors. Firstly, GPU memory capacity often does not align well with the power-of-2 nature of state vector sizes. Even when alignment is achieved, metadata storage limits complete memory utilization. Secondly, even with full GPU memory usage, CPU and storage memory (10$\times$ to 100$\times$ larger) remain largely unused.

\qtree utilizes unused memory to store intermediate states, enabling simulations of larger circuits beyond current noisy quantum circuit simulators' capabilities. Furthermore, the hybrid CPU-GPU compute node setup allows state vector movement between memory systems, driven by CPUs, while GPUs simulate subcircuits, reducing state copy overhead and overall execution time.

\noindent \textbf{3. Noise Modelling and Fidelity:} It is vital to model the impact of various noise on the output. To this end, \qtree supports a wide range of noise models. Our evaluations on diverse circuits show that \qtree can enable accurate simulations while supporting noise models such as depolarizing channels, thermal relaxation, and amplitude and phase-damping channels. Moreover, our experiments demonstrate a crucial balance between accuracy loss and speedup when reusing intermediate states. Notably, \qtree shows significant speedup with negligible and bounded accuracy loss.

We evaluate \qtree using 48 quantum circuits from 8 different circuit classes and three different computing platforms - \emph{Single Node CPUs, Single Node GPU, and CPU Cluster}. We show that \qtree achieves up to \maxSpeedup\ speedup over baseline noisy implementation (average speedup of \averageSpeedup). \qtree produces a result with a normalized fidelity that is within \maxFidelityDiff range of the baseline result.

\begin{figure*}[]
     \centering
    \subfloat[\centering (a) 3-qubit Bernstein-Vazirani circuit.]
    {\raisebox{5ex}%
    {\includegraphics[width=0.18\textwidth]{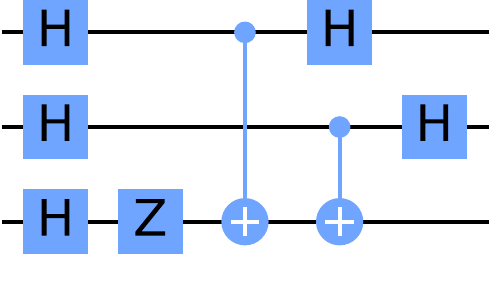} }
    \label{fig:bv-circ}}%
    \subfloat[\centering (b) Classical simulation of the 3-qubit Bernstein-Vazirani circuit.]
    {\raisebox{3ex}
    {\includegraphics[width=0.25\textwidth]{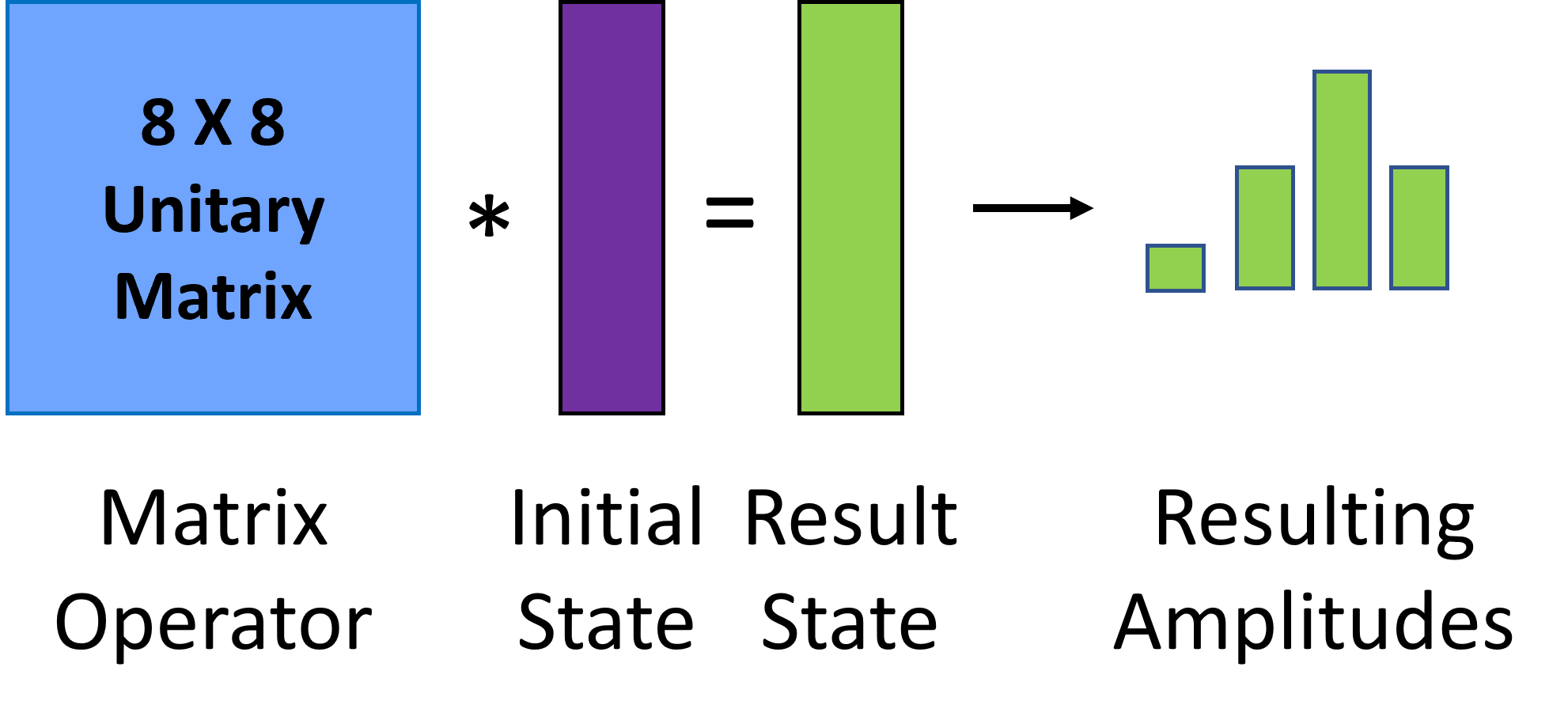} }
    \label{fig:bv-circ-comp}}%
     \subfloat[\centering (c) Noisy-version Bernstein-Vazirani circuit.]
    {\raisebox{7.5ex}
    {\includegraphics[width=0.26\textwidth]{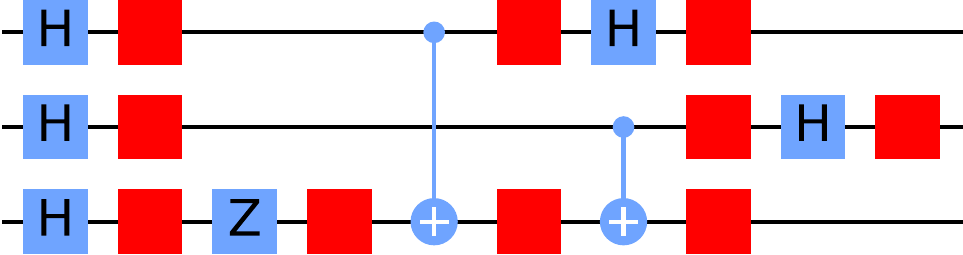}}
    \label{fig:noise-bv}}%
    \subfloat[\centering (d) Computations for the noisy simulation.]
    {{\includegraphics[width=0.24\textwidth]{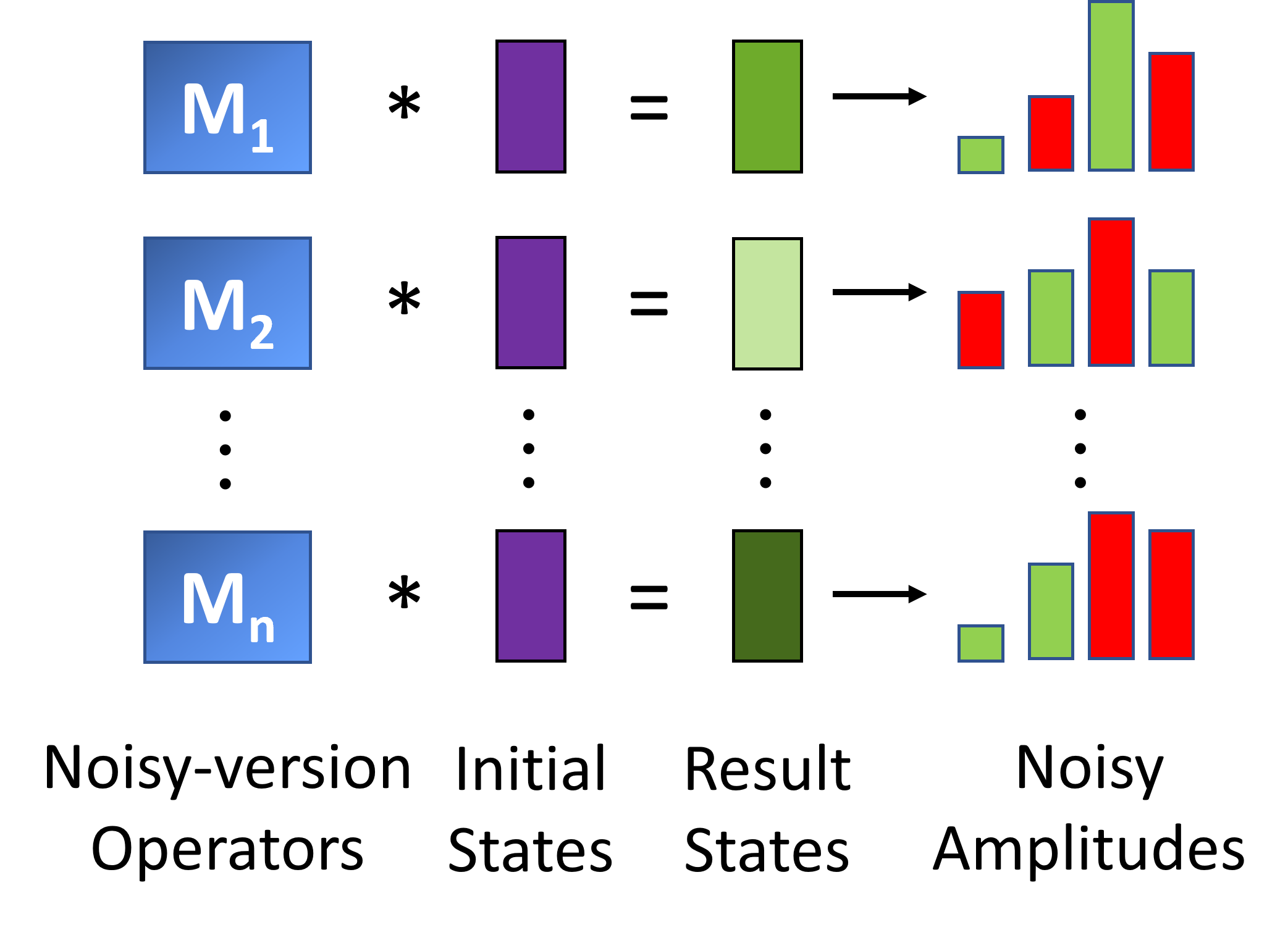} }
    \label{fig:noisy-sim-comp}}%
    \caption{Ideal and noisy simulation of the Bernstein-Vazirani circuit. (a) 3-qubit ideal circuit. (b) Ideal simulation. (c) Noisy circuit modeled with a depolarizing noise model. It should be noted that \qtree supports a wide range of noise models. The evaluation for these models is showcased in Section~\ref{noise_models_evaluation}. (d) Noisy simulation. }%
     \label{fig:bv_sim}%
\end{figure*}

\section{Background and Motivation}
\label{Background}

\subsection{Quantum Computing: Basics}
A quantum bit or qubit is the basic unit in quantum computing (QC). Its state, 
$\ket{\psi}$, can be expressed as:
\begin{equation}
 \ket{\psi} = \alpha \ket{0} + \beta \ket{1}
\end{equation}

Where $\ket{0}$ and $\ket{1}$ are orthogonal basis states, and $\alpha$ and $\beta$ are their probability amplitudes. For n-qubit systems, there are $2^n$ basis states, leading to 2$^n$ amplitudes. The qubit system state is usually represented as a state vector of amplitudes. Quantum algorithms are typically represented using circuits composed of ordered quantum gates. Quantum subcircuits are consecutive subsets of gate operations. Common gates include \emph{Pauli-X}, \emph{Pauli-Y}, \emph{Pauli-Z}, \emph{Hadamard (H)}, \emph{T}, and \emph{CNOT}. Circuit width indicates the qubit count, while circuit length denotes the number of gates in the circuit~\cite{wang2024optimizingftqcprogramsqec,qasmtrans}.

\begin{figure}[b]
    \centering
    \includegraphics[width=0.8\linewidth]{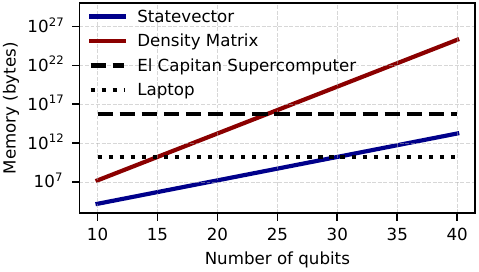}
    \caption{Memory overhead of density matrix vs. statevector simulators: On El Capitan, the world's top-1 supercomputer, the density matrix simulator handles fewer than 25 qubits, while the statevector simulator manages over 30 qubits on a personal laptop with 16GB of memory.}
    \label{fig:memory_overhead}
\end{figure}
\subsection{Ideal Quantum Circuit Simulation}
\label{label:statevector} 
An ideal quantum circuit simulator computes the final state vector by multiplying the matrix representations of gates with the initial state vector. This final state vector serves as the basis for outcome sampling. The Bernstein–Vazirani (\texttt{BV}) algorithm~\cite{bernsteinvazirani}, illustrated in Figure~\ref{fig:bv-circ}, embodies this simulation process. Figure~\ref{fig:bv-circ-comp} shows the computation within the simulation. Here, the unitary matrix is multiplied by an initial state vector, and a resulting state vector is produced, from which the outcome is sampled.

\subsection{Noisy Quantum Circuit Simulation}

Simulating noisy quantum circuits is essential for understanding quantum algorithms in realistic scenarios where decoherence and operational errors are inevitable. The conventional method employs density matrices to represent mixed quantum states and to incorporate noise effects through quantum channels~\cite{nielsen2010quantum,breuer2002theory}.

\subsubsection{Mixed-State Simulation Using Density Matrices}
\label{section:density_matrix_simulator}

In the density matrix formalism, the state of an $n$-qubit system is described by a $2^n \times 2^n$ complex matrix. This approach allows for a complete and accurate representation of mixed states and noise processes. However, a significant challenge arises due to the exponential scaling of memory requirements. Specifically, the memory overhead grows as $\mathcal{O}(4^n)$, which quickly becomes impractical for simulating larger quantum systems. As shown in Figure~\ref{fig:memory_overhead}, while a standard laptop can run statevector simulations for over 30 qubits, density matrix simulations are far more computationally intensive - even El Capitan~\cite{elcapitan2024}, one of the world's most powerful supercomputers, can only handle density matrices for fewer than 25 qubits.

\subsection{Pure-State Stochastic Simulation Execution}
\label{section:memory_underutilization}

To overcome the limitations of the density matrix approach, pure-state stochastic methods like the quantum trajectories method and the Monte Carlo wave-function method have been proposed~\cite{dalibard1992wave,molmer1996monte,plenio1998quantum}. In these methods, the mixed-state dynamics are approximated by averaging over an ensemble of pure-state shots. Each shot is subject to stochastic processes that model the noise. The ensemble process in the pure-state stochastic simulation involves integrating noise operators into the original circuit~\cite{decoherent_noise_gate}, as shown in Figure~\ref{fig:noise-bv}. This integration increases the complexity of the circuit and the corresponding computational workload, as shown in Figure~\ref{fig:noisy-sim-comp}.

\begin{figure}[]
    \centering
    \includegraphics[width=0.95\linewidth]{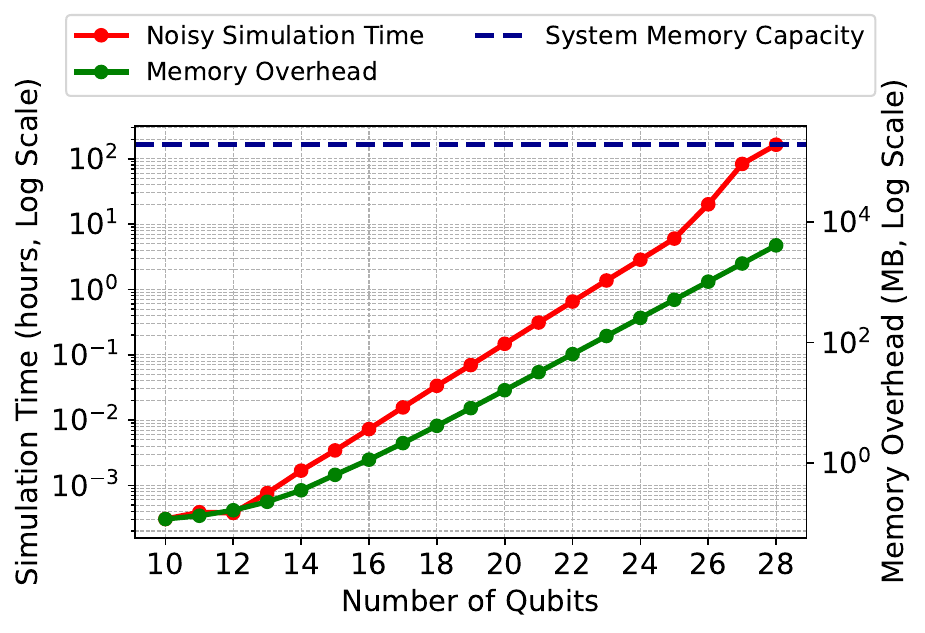}
    \caption{Simulation times and memory overhead for noisy \texttt{BV} circuits with 10 to 28 qubits. Each circuit involves 8192 shots and runs on dual 16-core Intel\textregistered\ Xeon\textregistered\ 6130 processors, having 192GB system memory. Both simulation time and memory overhead exhibit exponential growth. However, well before memory usage approaches system limits, noisy simulation times extend to hundreds of hours, establishing simulation time as the primary bottleneck for noisy tasks.}
    \label{fig:noisy_sim_mem_overhead}
\end{figure}

To evaluate the impact of this increased complexity, we measured the simulation times and memory overhead for \texttt{BV} circuits with 10 to 28 qubits using the density matrix approach (see Figure~\ref{fig:noisy_sim_mem_overhead}). Both simulation time and memory usage exhibit exponential growth. However, simulation times extend to hundreds of hours \emph{well before} memory usage approaches system limits. This indicates that -

\finding{\textit{While ideal quantum circuit simulation is typically memory-constrained, noisy circuit simulation is primarily limited by computational time, leaving memory resources under-utilized. This work aims to reduce computation time by leveraging these available memory resources to cache and reuse intermediate results.}}

By rethinking how memory is utilized in noisy quantum circuit simulations, we aim to improve their performance and address the computational bottleneck caused by the increased overhead.

\subsubsection{Error Analysis and Bounds}

The accuracy of stochastic simulations depends on the number of trajectories $N$ used in the ensemble. The statistical error decreases with the square root of $N$, following the central limit theorem~\cite{feller1991introduction}. The standard error $\epsilon$ is given by:
\begin{equation}
    \epsilon = \frac{\sigma}{\sqrt{N}}
\end{equation}
where $\sigma$ is the standard deviation of the observable across the ensemble. Consequently, the error bound can be controlled by adjusting $N$. Theoretically, as $N \rightarrow \infty$, the simulation results converge to those obtained from the density matrix approach~\cite{dalibard1992wave, molmer1996monte, plenio1998quantum,carmichael2009open}. This convergence is grounded in the equivalence between the ensemble average over stochastic pure-state evolutions and the solution of the master equation governing the density matrix dynamics.

\section{Tree-Based Quantum Simulator}
\label{Design}

\begin{figure*}[h!]
\centering
\includegraphics[width=\linewidth]{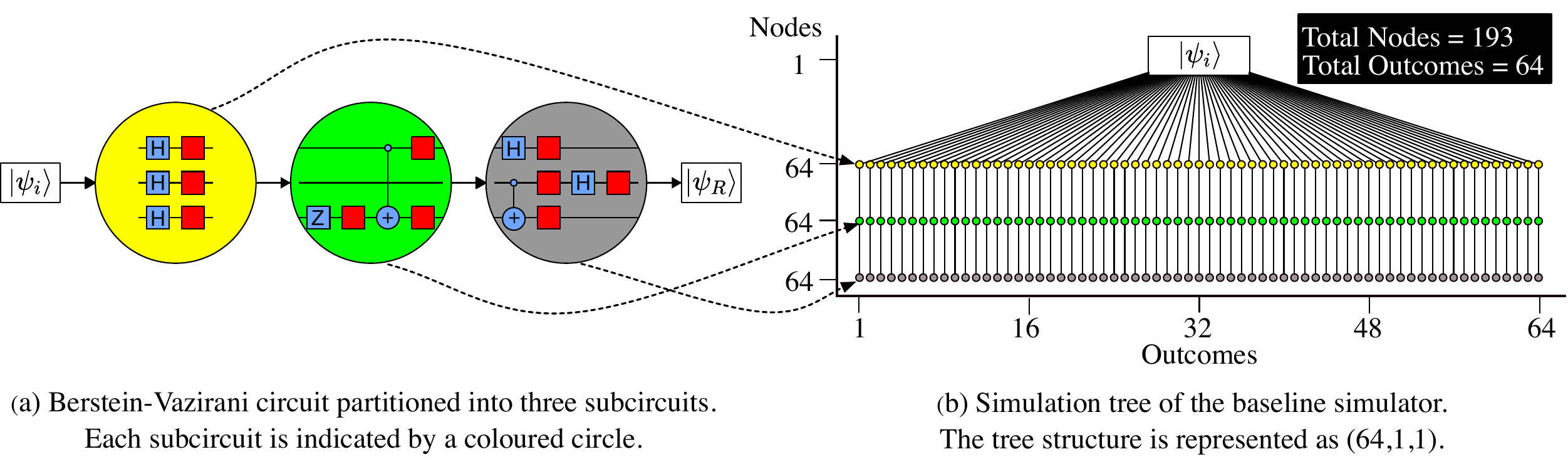}
\caption{Graphical representation of the Berstein-Vazirani circuit and the baseline simulation tree. The baseline simulation tree has 193 nodes -- 192 subcircuit nodes and one initial state node. The baseline simulation produces 64 outcomes. \label{fig:graphical-representation}}
\end{figure*}

This section presents the design of \qtree. \qtree uses a \emph{circuit partitioner} to divide the quantum circuit into subcircuits and determines the number of shots for each subcircuit. Figure~\ref{fig:graphical-representation}\textcolor{black}{a} shows the 3-subcircuit representation of the 3-qubit BV circuit. Figure~\ref{fig:graphical-representation}\textcolor{black}{b} shows the simulation tree with 64 shots when using the baseline statevector simulator. The nodes with a $(i+1)$ depth represent the $i_{th}$ subcircuit. The arity of the node indicates the number of reuses of the resulting state from that subcircuit. Thus, the arity of all the nodes in Figure~\ref{fig:graphical-representation}\textcolor{black}{b} is one except for the root node. 

\subsection{\qtree: Organization}
\qtree creates a simulation tree with nodes on the same layer, all having the same arity. We use the following notation to represent a given tree structure (assuming k subcircuits):
\[(A_0,A_1, ...,A_{k-1})\]
The $A_i$ is the arity of the nodes with a depth of $i$. For example, as the baseline simulation shown in Figure~\ref{fig:graphical-representation} incurs 64 shots, its simulation tree can be represented as (64,1,1). We can calculate the number of instances of i$_{\text{th}}$ subcircuit using the following equation: %
\begin{equation}
    \text{i$_{\text{th}}$ Subcircuit Instances} = \prod_{j=0}^{i-1}A_j
    \label{eq:circuit-instances}
\end{equation}
The total number of outcomes of a \qtree simulation tree is $\prod_{j=0}^{k - 1}A_j$.
\subsection{Dynamic Circuit Partition (DCP)}
\label{sec:dcp_design}
\subsubsection{Motivation}
\label{section:circuit_partition}

A straightforward technique, Uniform Circuit Partition (UCP), \emph{equally} divides the quantum circuit into k subcircuits with the same arity for all nodes. However, UCP's flaw lies in its exponential increase in the number of nodes as the tree depth increases. For instance, with three subcircuits and 1000 total shots, UCP yields a tree structure of (10,10,10). This results in the first subcircuit being simulated 10 times, the second subcircuit 100 times, and the third subcircuit 1000 times. This can lead to a result with high speedup, but potentially at the cost of accuracy.

We observe that the earlier nodes (i.e., the earlier parts of the circuit) are crucial for fidelity. To address the pitfalls of UCP, Exponential Circuit Partition (XCP) can be utilized, which assigns exponentially larger arities for earlier nodes, improving accuracy over UCP. However, XCP's constraint lies in its exponentially decreasing sequence of subcircuits. We present experimental results for UCP and XCP in Section~\ref{section:speedup_vs_accuracy}. To overcome UCP and XCP limitations, we propose Dynamic Circuit Partition (DCP), which dynamically determines the number of shots for the first subcircuit based on given error rates, aiming for speedup while preserving accuracy. DCP operates in two phases: circuit partitioning based on state copy overhead, followed by statistical determination of shot allocation.

\subsubsection{Generating First Subcircuit} 
DCP generates the first subcircuit based on state copy overhead [detailed in Section~\ref{state_copy_overhead}]. To ensure the benefits of state reuse outweigh the copy overhead, the subcircuit must exceed a minimum length. This minimum length is determined by balancing state copy cost against potential reuse benefits.

\subsubsection{Determining Shots for First Subcircuit}
The first subcircuit receives the fewest simulation instances. Thus, identifying the minimal nodes for the initial layer is crucial. We ensure near-optimal speedup and accuracy by creating the first subcircuit with the fewest gates, informed by state copy overhead considerations (Section~\ref{state_copy_overhead}). Its error rate is computed using Equation~\ref{eq:error_rate}:
\begin{equation}
    \text{First Subcircuit Error Rate} = 1 - \prod_i(1 - e_i)
    \label{eq:error_rate}
\end{equation}
where $e_i$ is the error rate of each gate in the initial subcircuit. This error rate guides the determination of node count for the first subcircuit, balancing accuracy and speedup. We adopt statistical sampling methods from the literature~\cite{sample_size} to allocate nodes effectively.

The idea is that \qtree selects a subset of the nodes (samples) from the baseline simulation tree (population) such that the selected samples can well-represent the original population. The sample size, therefore, is crucial to the final accuracy. The minimum number of nodes (sample size) is calculated using Equation~\ref{sample_size_formula} given below~\cite{sample_size}:
\begin{equation}\label{sample_size_formula}
A_0 \ge \frac{z^2*\hat{p}(1-\hat{p})}{\varepsilon^2}*\frac{1}{1+\frac{z^2*\hat{p}(1-\hat{p})}{\varepsilon^2N}}
\end{equation}

This equation calculates the minimum sample size to well-represent the population for a given confidence level ($z$) and margin-of-error ($\varepsilon$). Two additional parameters used in the equation are the total number of shots ($N$) and the overall error rate ($\hat{p}$) of the first subcircuit calculated using Equation~\ref{eq:error_rate}.

\subsubsection{Remaining Subcircuits}
With the initial subcircuit shots calculated for accuracy, the focus shifts to enhancing speedup. DCP divides the remaining circuit into 'k' subcircuits, each with an equal arity, resulting in a total of (k + 1) subcircuits. The equation $N = \prod_{j=0}^{k}A_j$ is used to determine arities for the remaining nodes, optimizing speedup. We use Equation~\ref{eq:subcirc_shots} for determining arities:

\begin{equation}
    A_1, ..., A_{k} = A_r = \text{floor}\left(\sqrt[k]{\frac{N}{A_0}}\right)
    \label{eq:subcirc_shots}
\end{equation}

\begin{figure}[h!]
    \centering
    \includegraphics[width=\columnwidth]{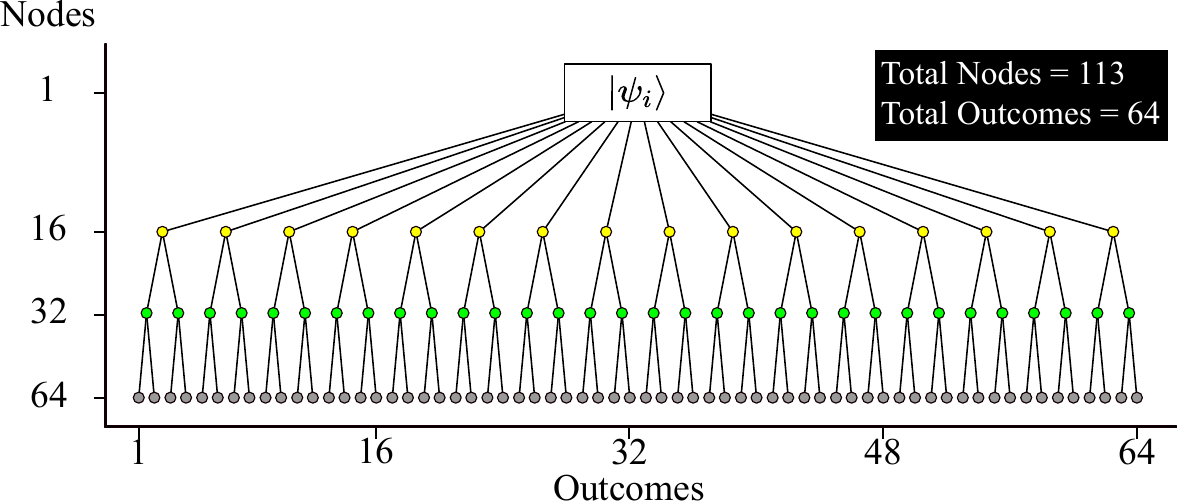}
    \caption{Simulation tree of \qtree using Dynamic Circuit Partition (DCP). The \qtree tree structure is (16,2,2). The simulation tree has 113 nodes and produces 64 outcomes.}
    \label{fig:dcp-64}

\end{figure}

A$r$ must be $\geq$2 for the remaining subcircuits to enable intermediate state reuse. DCP selects the maximum subcircuits (maximum `k') with A${r}$ $\geq$2, ensuring a minimum of user-specified outcomes. It then adjusts the uniform sequence, incrementing shots from the first subcircuit onward by one. This guarantees the desired outcomes. Next, DCP computes an alternative upper limit on subcircuits based on gate count and state copy overhead. The final number of subcircuits is set as the minimum of this value and the previously determined shot-based limit. The remaining circuit is then divided into equal-sized subcircuits based on the number of partitions determined. Figure~\ref{fig:dcp-64} illustrates this with a 3-qubit BV circuit simulation tree featuring three generated \qtree subcircuits.

\subsection{Memory Underutilization Problem}

Quantum circuit simulation uses a state vector with a length $2^n$, where n is the qubit count. Powerful High-Performance Computing (HPC) systems like Frontier~\cite{frontier2022}, Summit~\cite{summit2018}, and Perlmutter~\cite{perlmutter2022} are often used for large-scale circuit simulations due to their computational power. These systems, ranked among the top HPC systems worldwide, offer substantial computational capacity~\cite{top500}. Their massive parallel processing capabilities and high-bandwidth memory systems make them particularly well-suited for handling the intense computational demands of quantum circuit simulations. 

Besides computing, as shown in Table~\ref{tab:supercomputer_config}, these systems also offer copious amounts of memory. For instance, Frontier features nodes with 4x MI250X GPUs, each with 128GB of memory. However, due to the necessary storage for metadata and gate matrices, only 64GB per GPU can be utilized. Thus, the total available memory with 4 GPUs is 256GB. Similarly, Perlmutter uses 4x 40GB A100 GPUs per node, offering 160GB total memory with 4 GPUs. The maximum utilizable memory for quantum circuit simulation is 128GB. Summit, with 16GB GPUs and 6 GPUs per node, allows using 4 GPUs for balanced performance, resulting in a maximum utilization of 32GB.

\begin{table}[h!]
\caption{HPC System Configuration}
\centering
\resizebox{\columnwidth}{!}{%
\begin{tabular}{|l|l|l|l|}
\hline
System             & GPUs           & GPU    & CPU    \\
                   &                & Memory & Memory \\ \hline
Frontier (ORNL)    & 4x AMD MI250X  & 128 GB & 512 GB \\ \hline
Summit (ORNL)      & 6x NVIDIA V100 & 16 GB  & 512 GB \\ \hline
Perlmutter (NERSC) & 4x NVIDIA A100 & 40 GB  & 256 GB \\ \hline
\end{tabular}%
}
\label{tab:supercomputer_config}

\end{table}

Our experiments show that HPC systems can only utilize a fraction of their memory capacity for quantum circuit simulations: 25\% for Frontier, 5.3\% for Summit, and 30.8\% for Perlmutter. This significant underutilization of resources is evident.

\begin{figure}[b]
    \centering
    \includegraphics{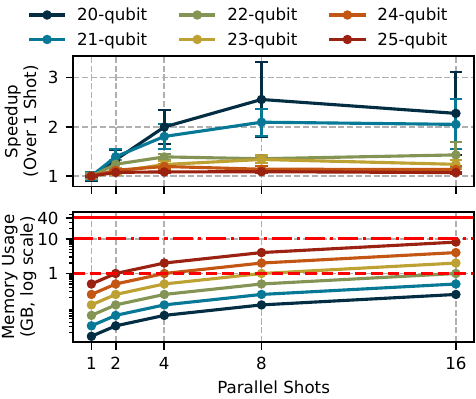}
    \caption{Performance analysis of parallel shot execution for a 1024-shot noisy QFT simulation (20–25 qubits) using Qiskit on an A100-40GB GPU. While circuits with 20–21 qubits gain up to a 3$\times$ speedup from parallel shots, the benefit diminishes with increasing qubit count. Beyond 24 qubits, parallel execution offers no advantage despite negligible memory usage (256MB per statevector, or 0.625\% of GPU memory).}
    \label{fig:embarassingly_parallel}
\end{figure}

The primary cause of memory underutilization is the computational overhead introduced by noisy simulation. The Monte Carlo process required for noisy quantum circuit simulation multiplies the baseline time complexity of $O(2^n)$ by a large constant factor, typically ranging from $2^{10}$ to $2^{17}$ repetitions. This effectively makes an $n$-qubit noisy simulation comparable in runtime to at least an $(n+10)$-qubit ideal simulation. For instance, a 30-qubit noisy simulation may require more computation time than a 40-qubit ideal simulation. This overhead cannot be mitigated through naive shot parallelization, as each shot fully utilizes the available computational resources. Figure~\ref{fig:embarassingly_parallel} presents experimental results for a 1024-shot noisy QFT simulation (20–25 qubits) with varying degrees of parallelism (2, 4, 8, and 16 parallel shots) on an A100-40GB GPU. The results show that while smaller circuits (20–21 qubits) benefit from parallel shot execution (up to 3$\times$ speedup), performance gains diminish as qubit count increases. Beyond 24 qubits, additional parallel shots provide no further speedup, even though each shot uses \emph{only} 256MB of memory (0.625\% of total available GPU memory).

\subsection{Intermediate States: Improve Utilization}

\begin{figure}[t]
    \centering
    \includegraphics[width=\linewidth]{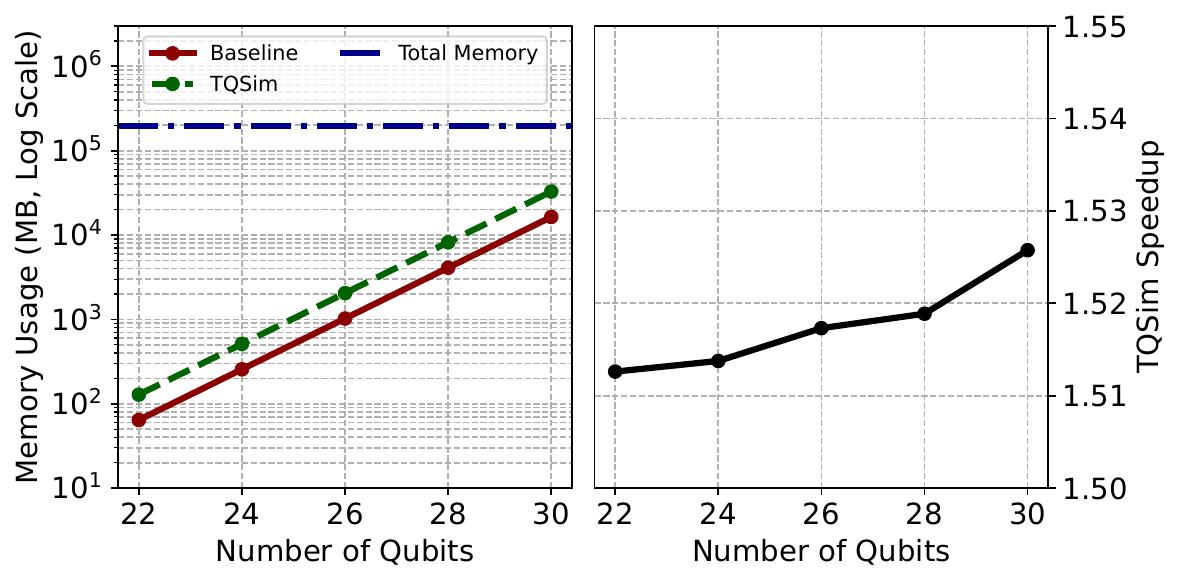}
    \caption{Simulation results of the Bernstein-Vazirani circuit with 22 to 30 qubits. The left figure illustrates the memory overhead of both the baseline and \qtree. While \qtree requires additional memory to store intermediate states, its peak memory consumption is still well below the system memory limit, represented by the green dashed line. The figure on the right shows the speedup \qtree achieves over the baseline. By leveraging the previously unused memory, \qtree attains a notable speedup compared to the baseline.}
    \label{fig:bv_memory_speedup}
\end{figure}

The design of \qtree solves a major underutilization problem. Currently, on the hardware platforms, it's generally hard to utilize most, if not all, system memories. This is particularly problematic for noisy quantum circuit simulations, as demonstrated in Section~\ref{section:memory_underutilization}, where the major bottleneck for scaling to larger quantum circuits remains the execution overhead, not the memory limitation. \qtree's design cleverly utilizes the unused memory in the baseline simulation to provide a significant speedup. Figure~\ref{fig:bv_memory_speedup} shows using \qtree to perform the noisy simulation of the \texttt{BV} circuit with 22 to 30 qubits. The figure on the right shows the memory usage of both \qtree and the baseline simulator. Despite \qtree taking additional memory to store intermediate states, \qtree's memory usage is still well below the system memory limit. In the meantime, \qtree achieves a significant speedup over the baseline without posing additional requirements on the underlying execution platform.

\subsection{Formulating the Error Bounds of \qtree}
\label{sec:error_bound}
Random samples are generated in baseline and \qtree simulations using the plug-in principle~\cite{pluginprinciple}. Let ${S_b}$ and ${S_t}$ represent the sets of state vectors produced by a layer of nodes in the baseline and \qtree simulations, respectively, with \emph{n} nodes in the baseline tree and \emph{m} nodes in the \qtree tree. Their difference is defined as~\cite{empiricalDistribution}:
\begin{equation} 
\label{eq:set_difference}
\begin{split}
  d(S_b,S_t) &= E[||S_b-S_t||^2]\\
    &= \sum_{i=1}^n \sum_{j=1}^m ||S_{b,i}-S_{t,j}||^2 %
\end{split}
\end{equation}
Here, $d(S_b,S_t)$ measures the average difference between the sets of state vectors $S_b$ and $S_t$, with $||S_n-S_m||$ representing the Euclidean norm of the difference. As \emph{m} increases, the difference decreases for a fixed \emph{n}. The first layer nodes in the simulation tree, with the largest difference in node count, are considered the worst-case scenario. Equation~\ref{sample_size_formula} is used to ensure that the difference between $S_b$ and $S_t$ is within the margin of error ($\varepsilon$) at a given confidence level ($z$).

Our experiment in Section~\ref{Evaluation} shows that the effect of noise is several orders of magnitude higher than the potential loss of accuracy caused by \qtree. Thus, the \qtree design ensures a high accuracy.
\subsection{Subcircuits Count vs State Copy Overhead}
\label{state_copy_overhead}

\begin{figure}[b]
    \centering
    \includegraphics[width=\columnwidth]{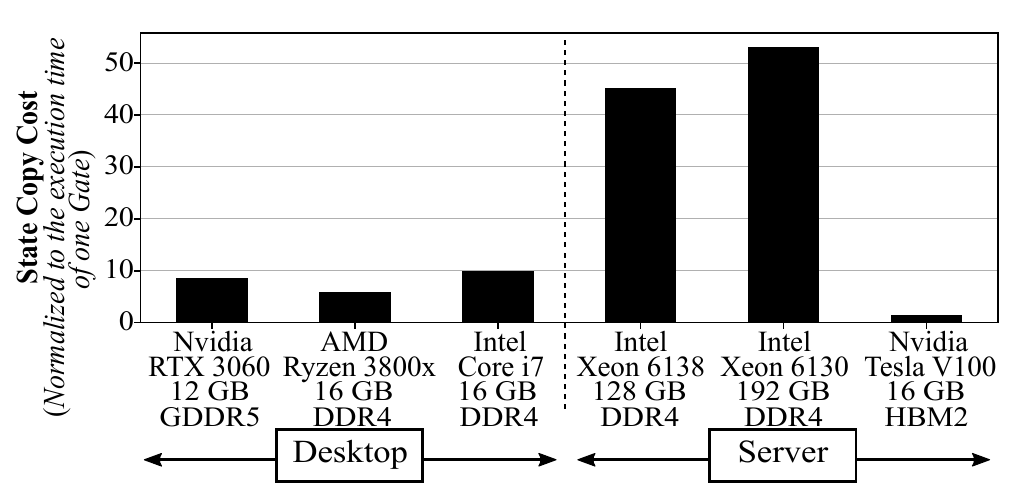}
    \caption{The state copy overhead across 3 CPU-based and 3 GPU-based hardware systems. The copying cost of the state vector is normalized to the execution time of one gate on the same machine. A cost of 20 gates means copying a state vector takes as long as executing 20 gates.}
    \label{fig:copy-overhead}
\end{figure}
The number of subcircuits determines the upper bound on the achievable speedup. For example, with two equal-length subcircuits, one can obtain a maximum speedup with a \qtree tree structure of (1, N). The corresponding maximum speedup is $\frac{1 + N}{2N}$ $\approx$ 1.5$\times$ and ignores the accuracy measure. We can easily achieve a similar speedup with a higher accuracy using more subcircuits.

The theoretical maximum speedup with `k' equal-length subcircuits is $\frac{kN}{(k-1) + N}$ where `N' is the number of shots. We can see that the maximum speedup increases as `k' increases. Therefore, a higher number of subcircuits results in a potentially higher speedup. However, we cannot naively increase the number of subcircuits as each additional subcircuit imposes a memory overhead (for storing the state) and execution overhead (for copying the state). To address the memory overhead constraint, we set the maximum number of subcircuits so that their size does not exceed the current memory capacity limit. To address the execution time overhead, we profile the state-copy cost using a set of profiling circuits and normalize it to the execution time of one gate. We use the state-copy cost to select the maximum number of subcircuits.

Figure~\ref{fig:copy-overhead} shows the state copy cost normalized to the execution time of one gate on the same machine for six systems. For example, for a desktop GPU, the time it takes to copy a state is approximately the execution time of 10 gates. The state copy cost is similar for circuits with 5 to 28 qubits. Therefore, we use an averaged state copy cost value for all circuit widths. We set the minimum number of gates in a subcircuit to equal the state copy cost. This way, the state copy overhead does not dominate the execution time. With this minimum number of gates limit, we effectively set the maximum number of subcircuits for a simulation task.

The state copy cost is much higher on server CPU systems. The reason is twofold. For one, the server memories run at 1.2$\times$ \emph{lower} frequency compared to desktop memories -- DDR4-3200 vs DDR4-2666, respectively. Therefore, it takes longer to copy a state on server systems. On the other hand, server systems have more high-performance cores than desktop systems and thus have a computing advantage. Thus, server systems take much less time to \emph{execute} a gate. These two combined factors result in a much higher \emph{normalized state copy cost} for server systems. Contrary to CPUs, the NVIDIA\textregistered\ Tesla\textregistered\  V100 system uses a faster HBM2 memory; thus, its state copy cost is also the lowest. The maximum number of sub-circuits in \qtree is determined such that we minimize the state copy and execution time overheads.

\section{Methodology}
\label{Methodology}
\subsection{Figure of Merit}
State fidelity is used as a metric to measure the similarity between two quantum states.
We use Equation~\ref{eq:fidelity} and Equation~\ref{eq:norm-fidelity}, as defined by Lubinski et al.~\cite{norm_fidelity}, to compute the state fidelity for noisy simulations. State fidelity is evaluated by computing the inner product of the state vectors between ideal and noisy results. State fidelity ranges from 0 to 1, where 1 indicates identical quantum states and 0 denotes completely different (orthogonal) states.
\begin{equation}
\displaystyle F_s(P_\text{ideal},P_\text{output})= \left( \sum_x \sqrt{P_\text{ideal}(x)P_\text{output}(x)}\right)^2
\label{eq:fidelity}
\end{equation}

One problem with the fidelity metric is that $F_s$ is not 0 when the output is completely random, i.e., the $P_\text{output}$ is uniform. To address this, we use the normalized fidelity metric, as defined by Hashim et al.~\cite{norm_fidelity1} and Lubinski et al.~\cite{norm_fidelity}, shown below.
\begin{equation}
\displaystyle F(P_\text{ideal},P_\text{output}) = \frac{F_s(P_\text{ideal},P_\text{output}) - F_s(P_\text{ideal},P_\text{uni})}{1 - F_s(P_\text{ideal},P_\text{uni}))}
    \label{eq:norm-fidelity}
\end{equation}
For simulators lacking noise modeling, accuracy is assessed by comparing their output similarity to an ideal reference simulator, as seen in prior studies~\cite{cutqc,full_state_compression}. However, directly applying normalized fidelity to \qtree is challenging due to the probabilistic errors in baseline noisy simulators. To accurately estimate the error in simulation, we compute a reference normalized fidelity using the baseline noisy simulator and compare it with \qtree's.

\subsection{Benchmarks}

We utilize quantum circuits ranging from 6 to 25 qubits from QasmBench, Qiskit, Cirq, and Qualcs. We include arithmetic operations like Adders, Multipliers, Quantum Fourier Transform (\texttt{QFT}), and Quantum Phase Estimation (\texttt{QPE}). Additionally, we employ near-term quantum algorithms such as Quantum Approximate Optimization Algorithm (\texttt{QAOA}) and Bernstein-Vazirani (\texttt{BV})~\cite{redqaoa,tomesh2022supermarq,smith2023clifford,informr}.

To assess the accuracy and speedup of \qtree, we employ Quantum Supremacy (\texttt{QSC}) and Quantum Volume (\texttt{QV}) circuits. These circuits lack structure, making them challenging to simulate. They are also used for benchmarking quantum hardware. For instance, \texttt{QV} circuits are run on both the simulator and real hardware to compute Quantum Volume, using the simulator output as a reference. Table~\ref{table:benchmarks} summarizes the key parameters for these circuits.

\begin{table}[]

\centering
\begin{small}
\caption{Benchmark Characteristics}
\setlength{\tabcolsep}{0.15cm} 
\renewcommand{\arraystretch}{1.3}
 \resizebox{0.8\columnwidth}{!}{
\begin{tabular}{|c|c|c|c|} 
 \hline
 Benchmark & Description & Width & Gate \\ 
          &  &  & Counts \\ 
 \hline\hline
 \texttt{ADDER} & Quantum Adder~\cite{qiskit2024, qasmbench}  & 4-10 & 16-133 \\ \hline
 \texttt{BV} & Bernstein-Vazirani~\cite{bernsteinvazirani, qiskit2024}  & 6-16 & 16-46 \\\hline
 \texttt{MUL} & \makecell{Quantum Multiplier~\cite{qiskit2024}}  & 13-25 & 92-1477 \\\hline
 \texttt{QAOA} & \makecell{Quantum Approx.\\ Optimization Algorithm~\cite{qaoa}}  & 6-15 & 58-175 \\\hline
 \texttt{QFT} &\makecell{Quantum Fourier Transform~\cite{qiskit2024}} & 10-20 & 237-975  \\\hline
 \texttt{QPE} &\makecell{ Quantum Phase Estimation~\cite{qiskit2024,qasmbench}} & 4-16 & 53-609 \\\hline
  \texttt{QSC} & \makecell{Quantum Supremacy Circuit~\cite{supremacy,cirq_developers_2020_4064322}}  & 8-16 & 38-160 \\\hline
 \texttt{QV} & Quantum Volume~\cite{quantum_volume,qulacs}   & 10-20 & 330-660 \\\hline
\end{tabular}
}
\label{table:benchmarks}
\end{small}
\end{table}

\noindent{\textbf{Why BV as a benchmark?}} BV relies on Clifford gates and can be efficiently simulated under Pauli noise using stabilizer simulations. However, the number of gates in BV increases linearly with the number of qubits, which means that when using the state vector simulation, the memory required to store quantum states scales much faster than computation, posing a challenge for TQSim. Moreover, BV's single-bit output makes it highly susceptible to simulation errors. Thus, \textit{BV is the worst-case benchmark for assessing TQSim's ability to balance accuracy and computational reuse}.

\subsection{Simulation Parameters}\label{section:params}
\noindent\textbf{1. Number of Shots:} We use 32,000 shots across various benchmarks, ensuring adequacy for the noisy quantum circuits with 6 to 25 qubits. Additionally, we conduct sensitivity tests by varying the number of shots to assess \qtree's accuracy and speedup. We perform two tests with reduced shot counts of 1000 and 3200 to magnify the noise impact.

\noindent\textbf{2. Noise Models:}
\label{section:noise_model}
We use the depolarizing noise model to highlight the benefits of \qtree. Additionally, for sensitivity studies, we verify the accuracy of \qtree with noise models that are constructed using various error channels, including:
\begin{itemize}[itemsep=0mm,leftmargin=*]
    \item \textbf{Depolarizing Channel (DC)}: Using Pauli operators (X, Y, Z)~\cite{nielsen_quantum_2012} to model noise.
    \item \textbf{Thermal Relaxation Channel (TR):} Modeling the decoherence using the T1, T2~\cite{abragam_principles_1961}, and gate times.
    \item \textbf{Amplitude Damping (AD):} Models energy relaxation of the qubit systems through a set of Kraus operators~\cite{nielsen_quantum_2012}. In our test, we use a damping ratio of 0.01.
    \item \textbf{Phase Damping (PD):} Phase damping channel also uses a set of Kraus~\cite{nielsen_quantum_2012} operators to model phase damping noise. We use a damping ratio of 0.01.
    \item \textbf{Readout (R):} During measurement, a measured classical bit is flipped with a given probability.
\end{itemize}

\noindent\textbf{3. Error Rate:} We use realistic device error rates obtained from Google Sycamore~\cite{quantum_datasheet, supremacy}. For error channels that do not have profiled device parameters, we select the conservative error parameters that cause a large noise effect on the result.

\subsection{System Configuration and Baselines}\label{section:conf}
We evaluate \qtree simulator on a platform with two Intel\textregistered\ Xeon\textregistered\ Gold 6130 processors @ 2.10 GHz, each having 16 physical cores with 192GB DDR4-2666 memory. The multi-node scaling experiments use multiple platforms with the same setup. We also evaluate the performance of \qtree on GPU-driven simulation setups using NVIDIA\textregistered\ V100 card with 16 GB of VRAM and A100 card with 40GB of VRAM. The baseline simulator and \qtree use all compute cores and threads available in the CPU and GPU.

We implement and evaluate \qtree across multiple state-of-the-art quantum circuit simulation frameworks. Our core implementation leverages \textit{Qulacs}\cite{qulacs}, a high-performance quantum circuit simulator, where we incorporate a depolarizing noise model. Using this primary framework, we conduct performance characterization on single-node architectures, including CPU performance evaluation [Section \ref{sec:single_node_perf}], GPU acceleration analysis [Section\ref{sec:gpu_perf}], and numerical accuracy assessment [Section \ref{sec:accuracy}].

To demonstrate the generality and portability of our approach, we extend the implementation to several other prominent frameworks. We employ \textit{qHiPSTER}\cite{qHiPSTER} for distributed memory systems evaluation on multi-node CPU clusters [Section \ref{sec:multi_node}], \textit{Qiskit}\cite{qiskit2024} for verification under various quantum noise models [Section \ref{noise_models_evaluation}], and NVIDIA's \textit{CuQuantum} framework\cite{cuquantum} (specifically using CuStateVec 1.7.0) for GPU-accelerated simulation [Section \ref{sec:gpu_perf}]. This comprehensive evaluation methodology enables us to assess \qtree's performance, accuracy, and scalability across diverse computing platforms and simulation environments.

\section{Evaluations: Single and Multi Node}
\begin{figure*}
    \centering
    \begin{minipage}[t!]{.78\linewidth}
    
    \begin{center}
        \includegraphics[width=0.4\linewidth]{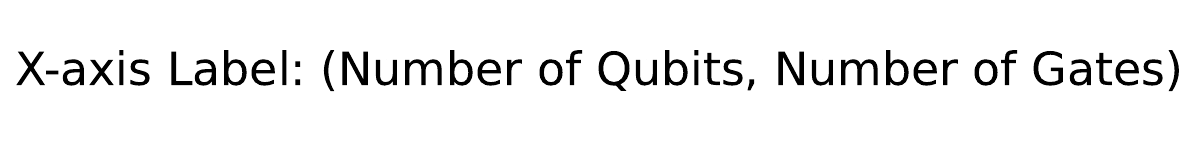}
    \end{center}  
      
    \vspace{-0.2in}
    
    \subfloat[\centering (a) \texttt{ADDER} - 2.20$\times$]
    {{\includegraphics[width=0.21\linewidth]{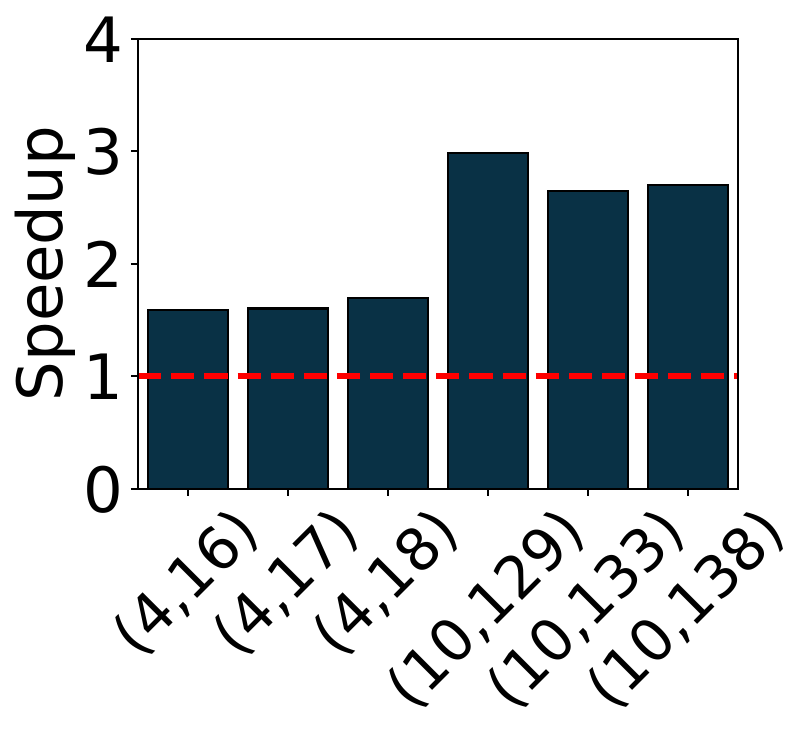}}
    \label{speedup:adder}}%
    \subfloat[\centering (b) \texttt{BV} - 1.77$\times$]
    {{\includegraphics[width=0.21\linewidth]{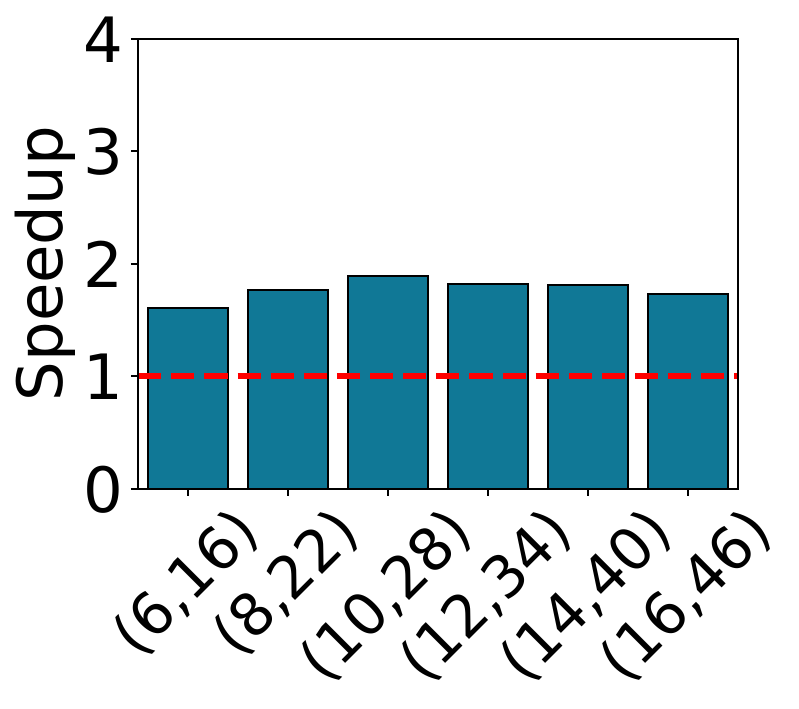}}
    \label{speedup:bv}}%
    \subfloat[\centering (c) \texttt{MUL} - 2.62$\times$]
    {{\includegraphics[width=0.205\linewidth]{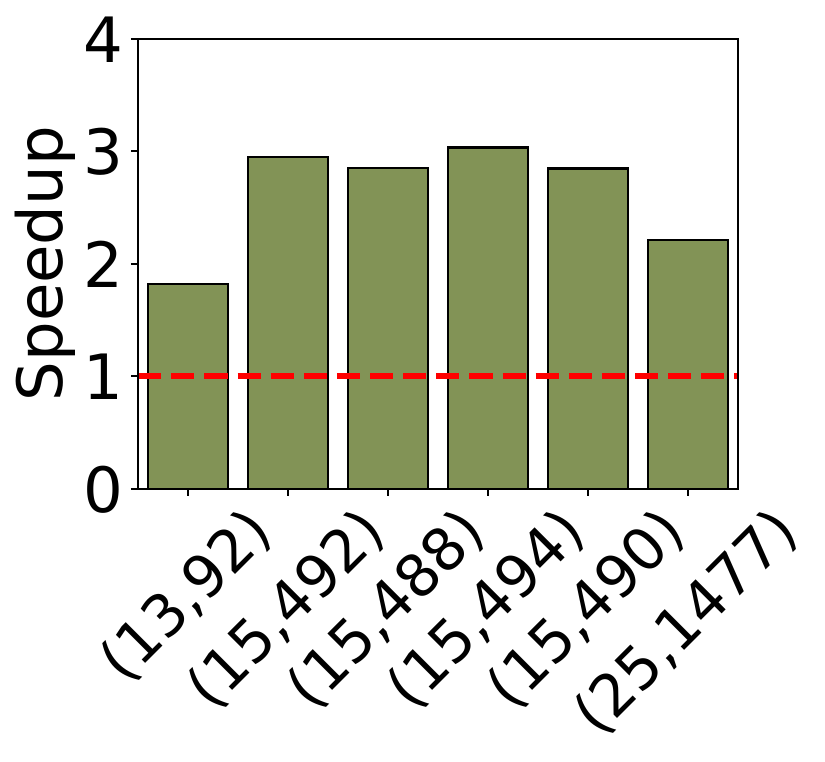} }
    \label{speedup:mul}}%
    \subfloat[\centering (d) \texttt{QAOA} - 2.39$\times$]
    {{\includegraphics[width=0.205\linewidth]{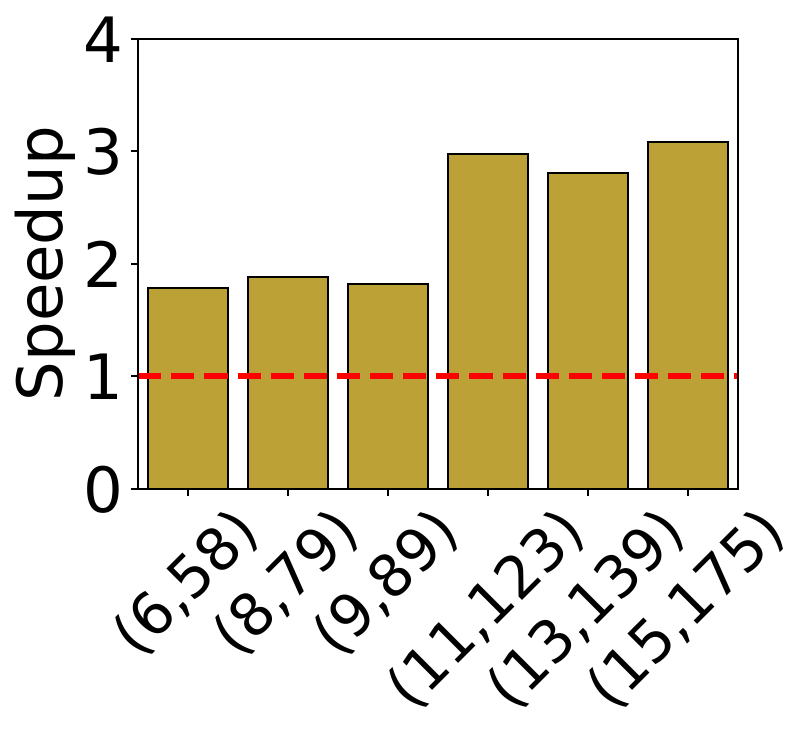}}
    \label{speedup:qaoa}}%

    \subfloat[\centering (e) \texttt{QFT} - 3.10$\times$]
    {{\includegraphics[width=0.21\linewidth]{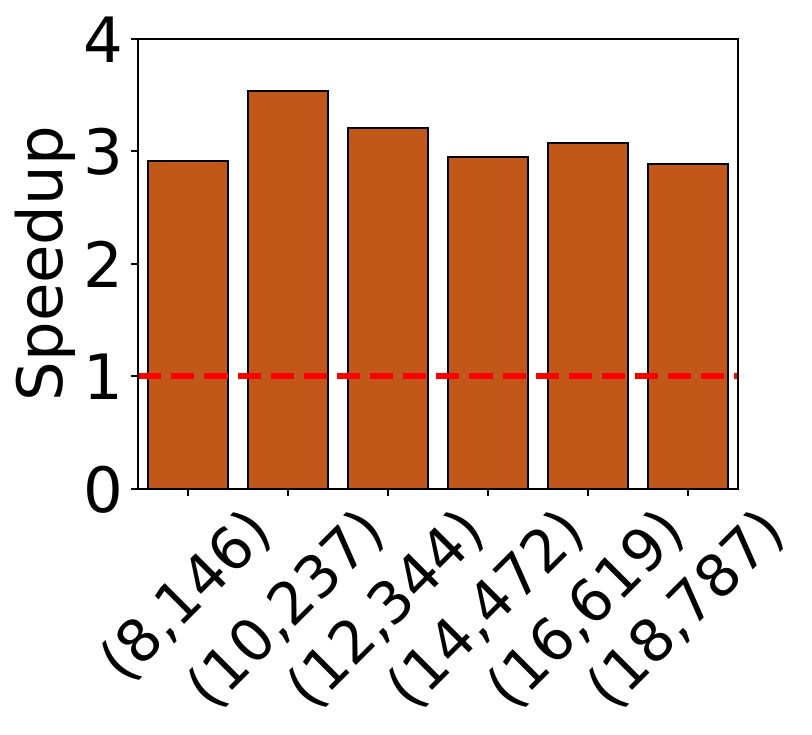}}
    \label{speedup:qft}}%
    \subfloat[\centering (f) \texttt{QPE} - 2.76$\times$]
    {{\includegraphics[width=0.21\linewidth]{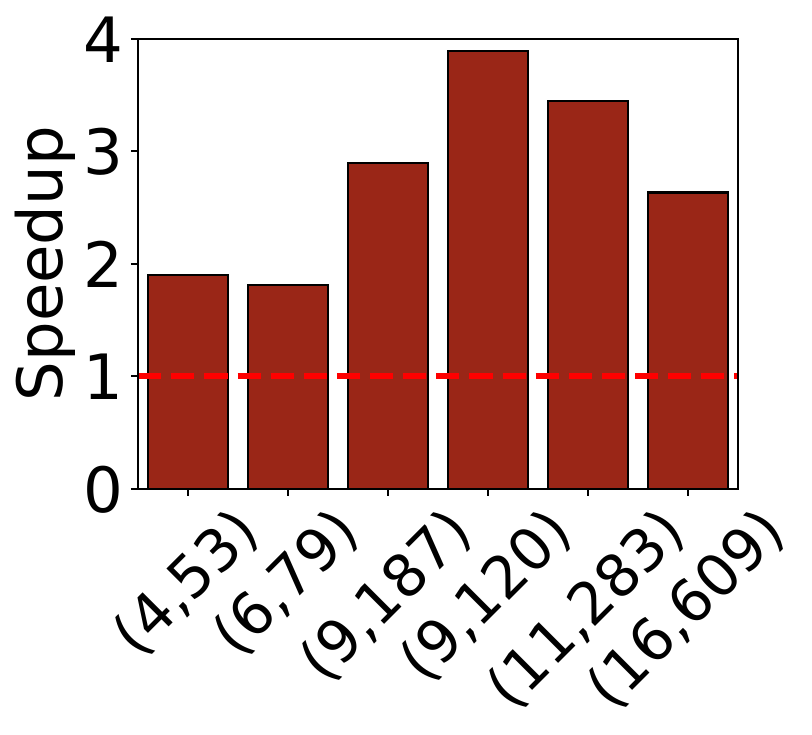} }
    \label{speedup:qpe}}%
    \subfloat[\centering (g) \texttt{QSC}  - 2.22$\times$]
    {{\includegraphics[width=0.205\linewidth]{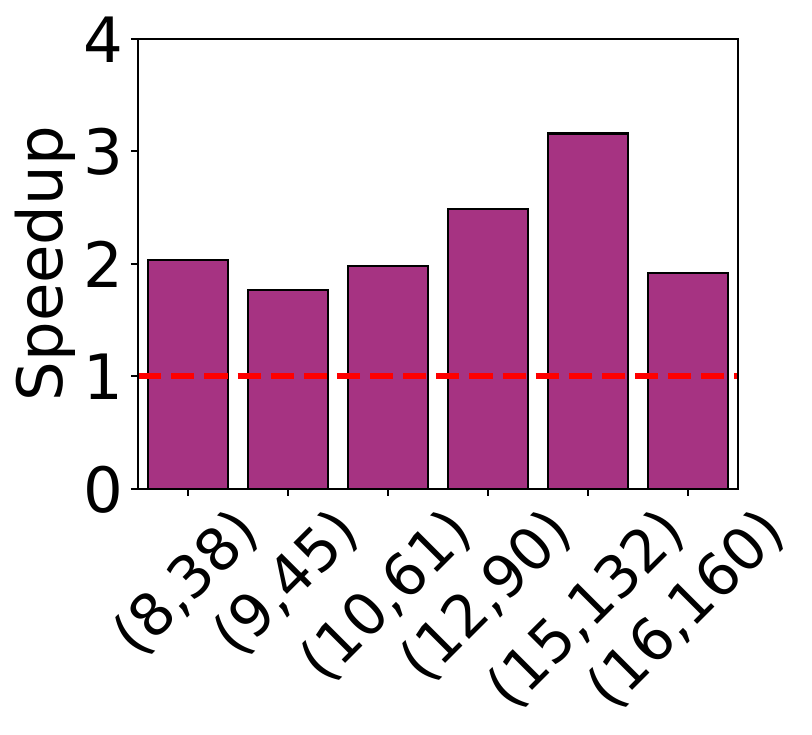}}
     \label{speedup:qsc}}%
   \subfloat[\centering (h) \texttt{QV} - 2.98$\times$]
    {{\includegraphics[width=0.205\linewidth]{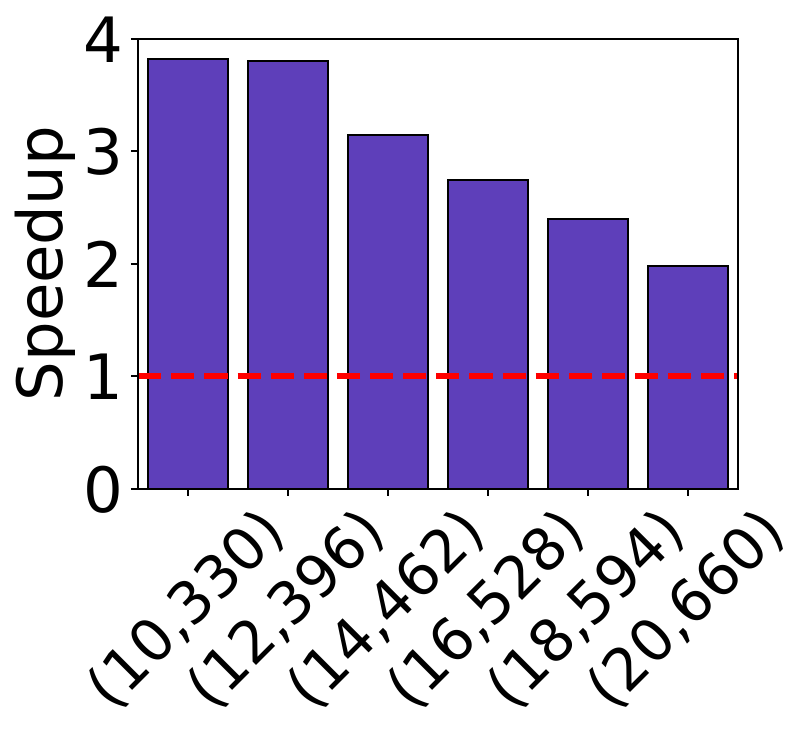}}
    \label{speedup:qv}}%
    \end{minipage}
    \begin{minipage}[t]{.20\linewidth}
    \vspace{-0.85in}
    \subfloat[\hspace{-0.85in} (i) \qtree Performance]
    {{\hspace{-0.85in}\includegraphics[width=1.8\linewidth]{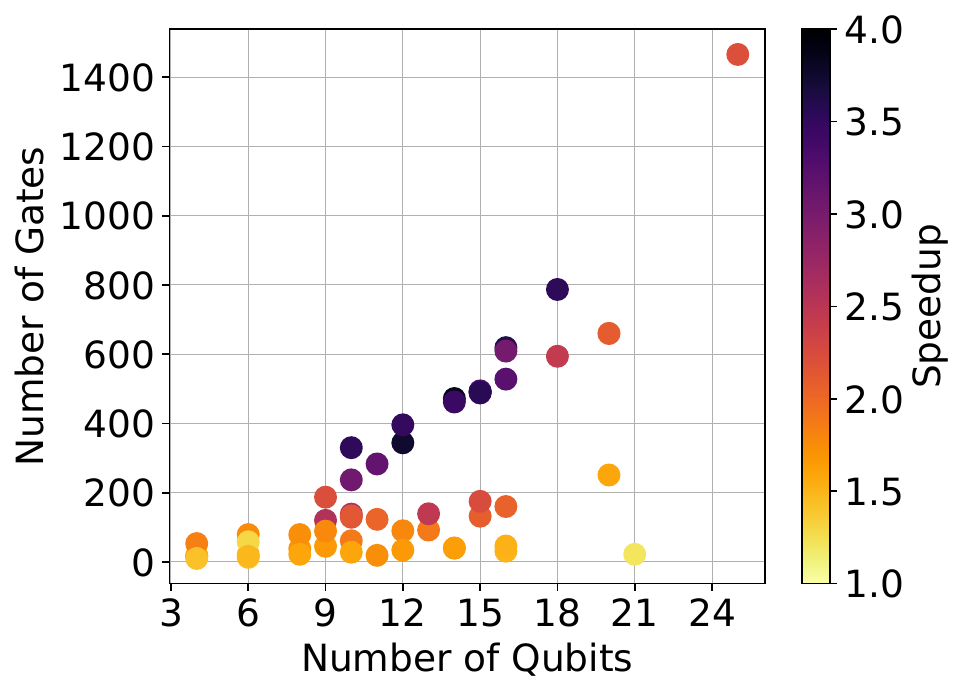}}}
    \label{speedup:heatmap}
    \end{minipage}
    
    \caption{Speedups of \qtree over the baseline simulator for 8 benchmark circuits. The tuple indicates the number of qubits and gates of the circuit. For example, the first ADDER circuit in (a) has 4 qubits and 16 gates.}%
    \label{fig:speedups}%
\end{figure*}

\label{Evaluation}

\qtree leverages computation reuse in noisy quantum circuit simulations. We evaluate the performance of \qtree on three platforms - Single CPU, GPU, and multi-node CPU cluster. Furthermore, we test the accuracy of \qtree with frequently used noise models and simulation configurations described in Section~\ref{section:params} and Section~\ref{section:conf}.

\subsection{Performance on Single CPU Node}
\label{sec:single_node_perf}

Figure~\ref{fig:speedups} shows the speedup of \qtree over the baseline \textit{Qulacs} simulator for all the benchmark circuits. Overall, \qtree is \minSpeedup to \maxSpeedup faster compared to the baseline simulator, and on average, it provides \averageSpeedup speedup. In general, shallow circuits, such as \texttt{ADDER}, have the least room for improvement as the number of subcircuits is limited. \qtree still achieves 2.2$\times$ speedup for such circuits.

We can increase the number of subcircuits with increasing circuit lengths to enable higher speedup. However, to maintain a high simulation accuracy for the realistic error rates, \qtree limits the maximum number of subcircuits. For example, to simulate \texttt{QFT\_14} with 472 gates with a 0.1\% gate error rate, \qtree partitions the input circuit into seven subcircuits and assigns 500 shots to the first subcircuit. This results in a theoretical maximum speedup of 3.53$\times$. As shown in Figure~\ref{speedup:qft}, \qtree provides 3.21$\times$ speedup for \texttt{QFT\_14}, close to the theoretical maximum speedup. This indicates that the benefits of circuit partition and state reuse are higher than the overhead in creating copies for the subcircuit.

Circuits with large widths and short lengths cannot be partitioned into multiple subcircuits. Also, the intermediate state transfer overhead for the high-width subcircuit is significantly higher than for circuits with a smaller width. These two factors result in lower speedups for such benchmarks. For example, the \texttt{BV} circuits in Figure~\ref{speedup:bv} can only be partitioned into two subcircuits. The resulting average speedup for \texttt{BV} circuits is 1.77$\times$. Table~\ref{table:simulation-time} shows the simulation time of three circuits with 18 to 20 qubits, showcasing the potential time savings of \qtree for medium-scale circuits. 

\begin{table}[]
\centering
\caption{Simulation Time: Medium-Scale Circuits}
 \resizebox{0.7\columnwidth}{!}{
\begin{tabular}{|c|c|c|c|}
 \hline
 Benchmark & \makecell{Baseline\\ Simulation\\ Time (s)} & \makecell{\qtree\\ Simulation\\ Time (s)}& Speedup \\ 
\hline\hline
\texttt{QV\_18} &  708.7 & 295.1 & 2.41$\times$ \\ \hline
\texttt{QV\_20} &  2123.5 & 1070.5 & 1.98$\times$ \\ \hline
\texttt{QFT\_20} &  2783.8  & 963.8 & 2.89$\times$ \\ \hline
\end{tabular}}
\label{table:simulation-time}
\end{table}

Figure~\ref{fig:speedups} shows speedups of \qtree for benchmark circuits. Notably, \qtree faces its primary challenge in achieving significant speedups for square circuits with high width and short length, despite their high fidelity on NISQ hardware. The real hurdle arises when studying noise effects on longer circuits, where error susceptibility increases substantially. Simulating such circuits with parametric noise can enhance our understanding of the impact of noise on fidelity. However, conventional quantum simulators struggle with the slow simulation of high-length circuits. For instance, simulating the 20-qubit \texttt{QFT} circuit for 32,000 shots takes approximately 46 minutes with the baseline simulator, whereas \qtree achieves a speedup of 2.89$\times$ compared to baseline.

\subsection{Performance on Single GPU Node}
\label{sec:gpu_perf}

\begin{figure}[b]
    \centering
    \includegraphics{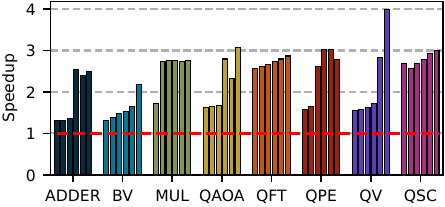}
    \caption{\qtree speedup over CuQuantum baseline, achieving 2.3$\times$ average and up to 3.98$\times$ across benchmarks.}
    \label{fig:cuquantum}
\end{figure}

Figure~\ref{fig:cuquantum} demonstrates the performance comparison between \qtree and the baseline GPU-based \textit{CuQuantum}~\cite{cuquantum} simulator. We utilize the CuStateVec (1.7.0) library's C/C++ APIs for this time-sensitive evaluation. \qtree achieves consistent speedup with CuStateVec (2.3$\times$ average, up to 3.98$\times$), comparable to its Qulacs-based CPU implementation. This consistency demonstrates that \qtree's performance improvements are from fundamental computation reduction rather than backend-specific optimizations.

\begin{figure}[b]
    \centering
    \subfloat[\centering (a) Strong Scaling.]
    {{\includegraphics[width=0.8\linewidth]{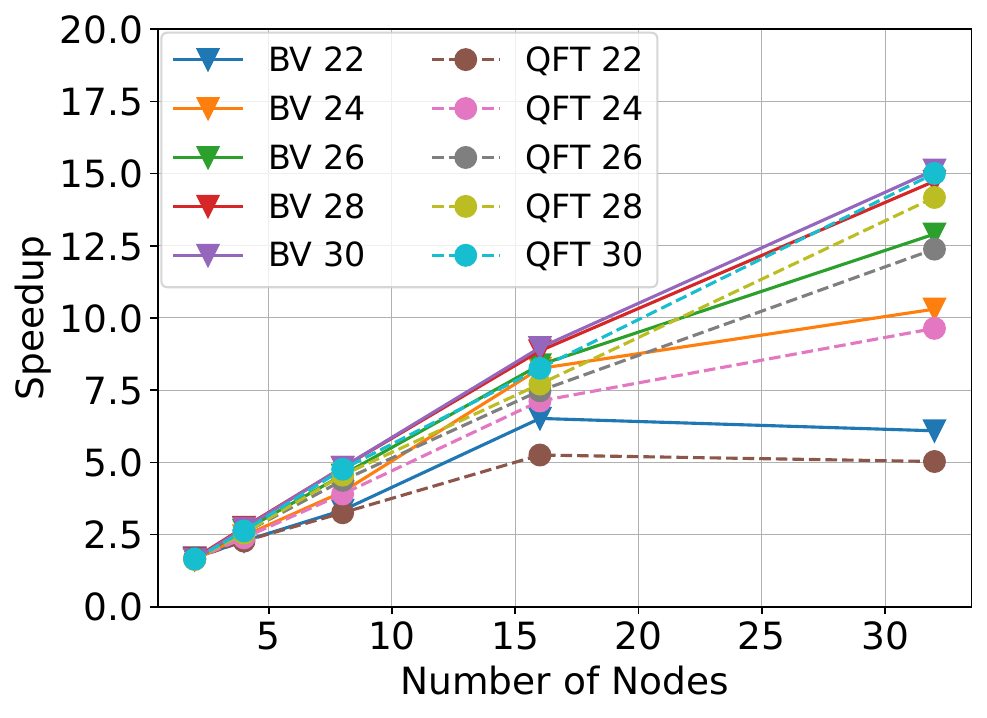} }
     \label{fig:strong_scaling}}%
     
    \subfloat[\centering (b) Weak Scaling.]
    {{\includegraphics[width=0.8\linewidth]{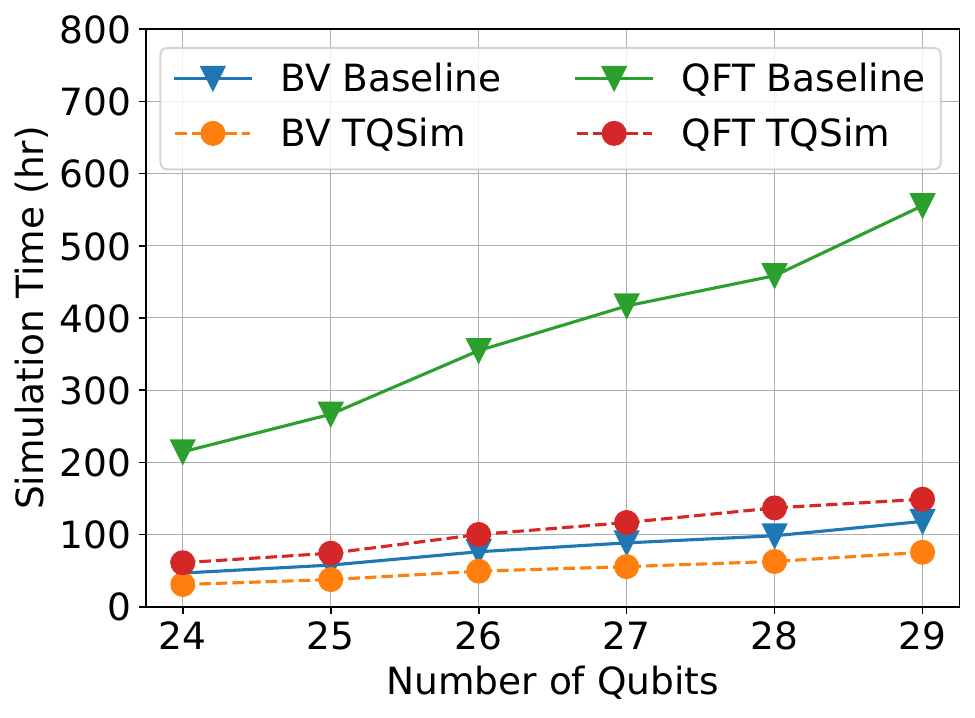} }
     \label{fig:weak_scaling}}%

    \caption{Strong and weak scaling of \qtree. (a) Strong scaling: Simulation time when varying the number of nodes. (b) Weak scaling: Simulation time when the computation per processor remains constant. The numbers of nodes used for 24 to 29 qubits are 1, 2, 4, 8, 16, and 32, respectively.}%
    \label{fig:gpu_and_scability}%
\end{figure}

Notably, \qtree's execution scheduler, originally designed with Qulacs as the simulator backend, requires only minimal modification to work with CuStateVec or other potential backends - simply replacing the underlying simulation functions. This dual advantage of \qtree – consistent performance gains across backends and straightforward backend integration – makes it a practical choice for quantum circuit simulation. Users can easily adapt their preferred simulation backend while maintaining the substantial speedups demonstrated in this paper.

Our results show that \qtree's performance on GPU aligns closely with its single-node CPU counterpart, primarily due to the preservation of the original computation task's parallel structure. It's worth noting that many existing optimization techniques for quantum circuit simulations, including multiprocessing and GPU acceleration, are predominantly designed for single-shot simulations. In the context of GPU implementation, the main additional overhead introduced by \qtree is the state copy operation on the GPU. However, our results indicate that this overhead is negligible when compared to the total simulation time, highlighting the efficiency of our approach even in GPU-accelerated environments.

\subsection{Performance on Multi-node CPU Cluster}
\label{sec:multi_node}

When simulating quantum circuits across multiple computing nodes, the state vector is equally partitioned and stored. To evaluate \qtree's performance in multi-node simulation, we implement TQSim based on qHiPSTER~\cite{qHiPSTER}, a multi-node quantum circuit simulator. We select \texttt{BV} and \texttt{QFT} circuits for evaluation, representing short and long circuits with one and seven intermediate states, respectively. Despite our infrastructure's ability to store selected circuit state vectors on a single node, we distribute them across multiple nodes to assess \qtree's additional state movement overhead in a hypothetical 'memory-constrained' multi-node scenario, where inter-node communications are necessary.

Figure~\ref{fig:strong_scaling} shows strong scaling results for benchmark circuits with 24 to 32 qubits across 1 to 32 nodes. Smaller circuits exhibit poor scaling due to limited computation and dominant communication overheads. Larger circuits, on the other hand, show improved scaling. The overall strong scaling performance of \qtree is closely aligned with the baseline qHiPSTER simulator.

To assess weak scaling, we conducted experiments using circuits ranging from 24 to 29 qubits, distributing the workload across 1 to 32 nodes while maintaining a constant computational load per gate per node. As depicted in Figure~\ref{fig:weak_scaling}, both the baseline simulator and \qtree experience performance impacts due to frequent inter-node communication. However, \qtree consistently outperforms the baseline across all test scenarios, demonstrating a significant speedup. Our analysis also reveals that for \texttt{QFT} and \texttt{BV} circuits, the simulation time increases proportionally with the gate count as the number of qubits grows. This relationship underscores the scalability of our approach in handling larger quantum circuits while maintaining performance advantages.

\subsection{Outcome Accuracy of \qtree}
\label{sec:accuracy}

\begin{figure}[t]%
   \centering
   \includegraphics[width=0.8\linewidth]{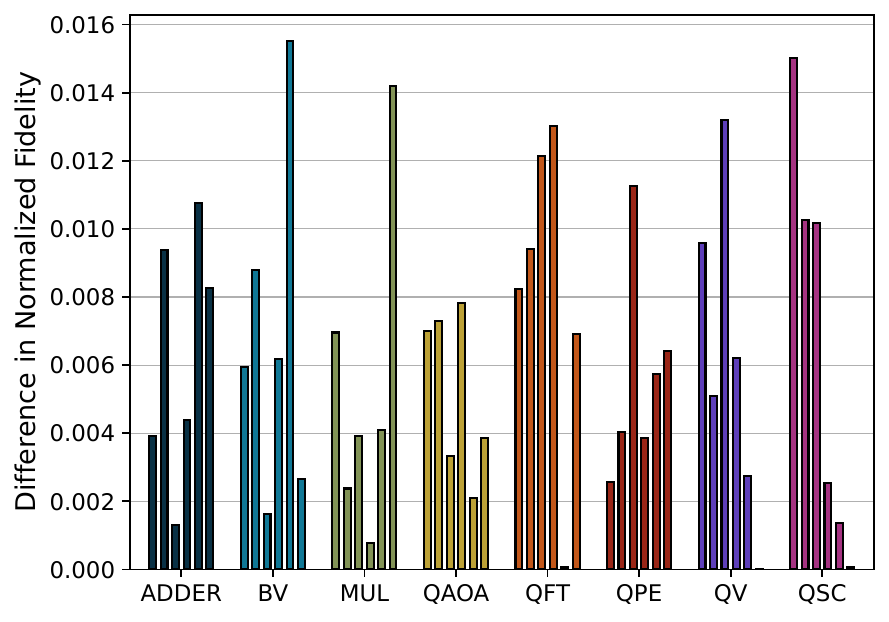}
   \caption{Baseline and \qtree normalized fidelity difference across 48 benchmarks. The average and maximum differences are only \averageFidelityDiff and \maxFidelityDiff (negligible), respectively.}%
   \label{fig:norm-fidelity-diff}%
\end{figure}
\begin{figure}[b]
    \centering
    \includegraphics[width=0.95\linewidth]{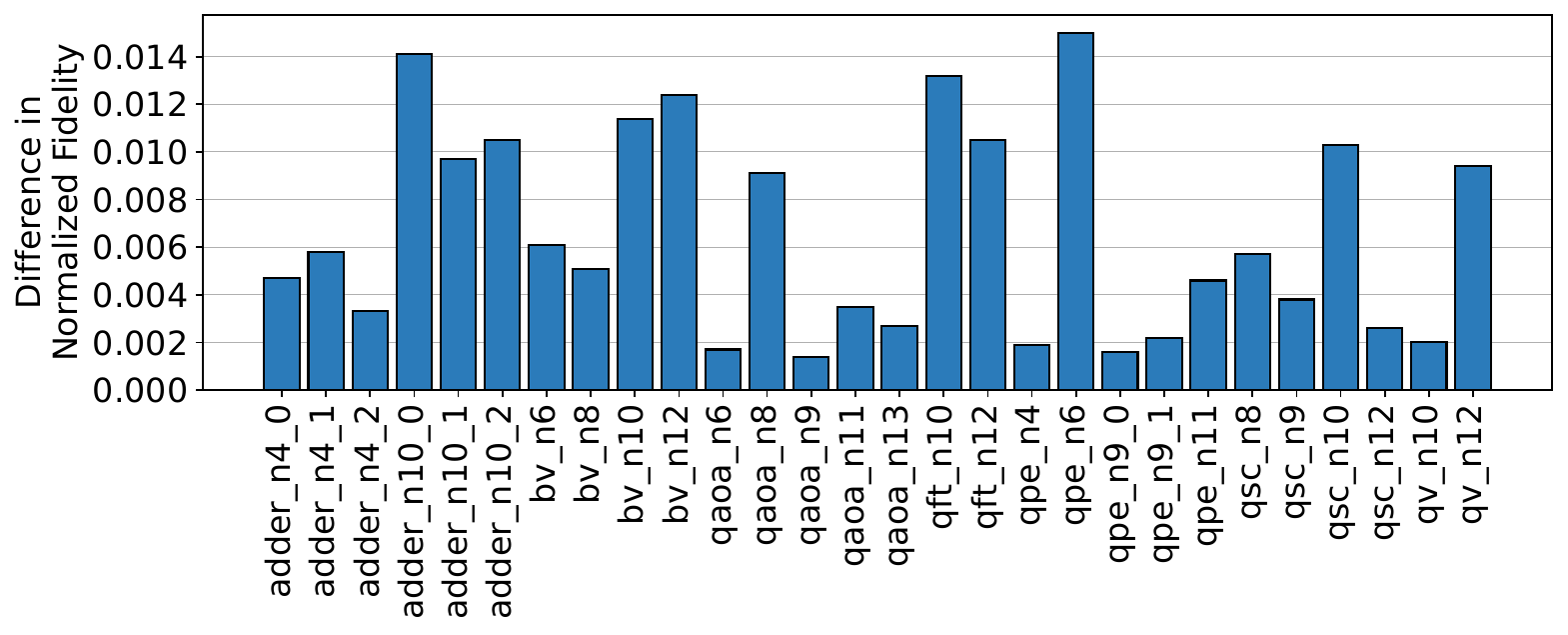}
    \caption{Difference in normalized fidelity of \qtree result compared to baseline density matrix simulation result.}
    \label{fig:dm_accuracy}
\end{figure}
Figure~\ref{fig:norm-fidelity-diff} compares the normalized fidelity for baseline and \qtree. We observe a less than \maxFidelityDiff difference in normalized fidelity values for all benchmarks. The benchmark circuits cover a wide range of circuit widths, lengths, and types of output distributions. Thus, with a fixed number of shots, we observe a wide range of baseline normalized fidelity. Still, \qtree provides a result with a normalized fidelity that is very close to the normalized fidelity of the baseline result. We provide a more in-depth analysis of the accuracy of the \qtree in Section~\ref{noise_models_evaluation}.

Figure~\ref{fig:dm_accuracy} shows the difference in normalized fidelity between the results of the \qtree and the density matrix simulators. As discussed in Section~\ref{section:density_matrix_simulator}, the density matrix simulator requires 2$^n\times$ more memory capacity, and we use feasible circuits for comparison. We observe an average difference in normalized fidelity of 0.007 and a maximum difference of 0.015. This is very similar to the result from the baseline state vector simulator.

\subsection{Sensitivity: Varying Noise Models}
\label{noise_models_evaluation}
We evaluate \qtree's accuracy using three circuits with diverse characteristics and output distributions. Each experiment is conducted 10 times, with the average normalized fidelity reported. Our baseline simulator results indicate that the depolarizing channel has the most significant impact on accuracy. Consequently, we use the depolarizing channel parameters to generate the \qtree structure, applying this approach across all noise model experiments.

\begin{figure}[t]
    \centering
    \includegraphics[width=0.8\linewidth]{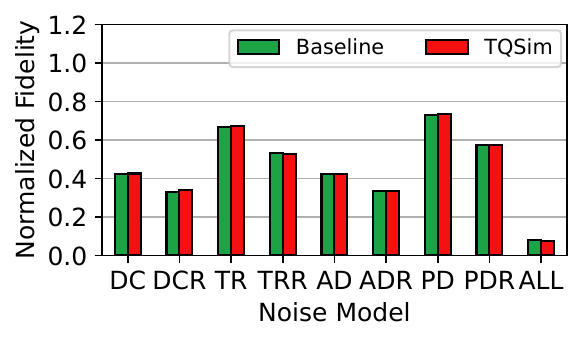} 
    \caption{Normalized fidelity values for the \texttt{QPE\_9} circuit executed with nine different noise models. Details of the error channels are provided in Section~\ref{section:noise_model}.}
    \label{fig:noise_model}
\end{figure}

Our analysis incorporates various noise models, including Depolarizing (DC), Thermal Relaxation (TR), Amplitude Damping (AD), Phase Damping (PD), and Readout (R). Figure~\ref{fig:noise_model} showcases the impact of different noise model combinations on normalized fidelity. For instance, \texttt{TRR} represents the application of both Thermal Relaxation and Readout noise. The 9-qubit \texttt{QPE} circuit, due to its high gate count, exhibits particular sensitivity to DC, TR, and AD. Despite this, \qtree achieves fidelity levels matching the baseline across all nine noise models tested. The 9-qubit \texttt{QPE} circuit is designed to estimate an eigenvalue that cannot be precisely represented by a 9-bit fixed-point number, resulting in a narrow bell curve output distribution. As illustrated in Figure~\ref{fig:noise_model}, \texttt{QPE\_9} demonstrates heightened sensitivity to noise, especially DC, TR, and AD. Nonetheless, \qtree consistently produces results that closely align with the baseline across all nine noise models, underscoring its robustness and accuracy in simulating noisy quantum circuits.

\subsection{Accuracy-Speedup Trade-off Analysis}\label{section:speedup_vs_accuracy}

We analyze the relationship between computational speedup and output accuracy using a 9-qubit \texttt{QPE} circuit containing 120 gates, with experiments conducted using 1000 shots. In addition to \qtree's DCP structure (250-2-2), we evaluate several alternative partitioning strategies: XCP (20-10-5) and UCP (10-10-10) as discussed in Section~\ref{section:circuit_partition}, two manually created structures (5-10-20, 2-2-250) designed for lower computational overhead, and an extreme case (250-1-1) that produces only \texttt{$A_0$} outcomes.

\begin{figure}[h!]
    \centering
    \includegraphics[width=0.8\linewidth]{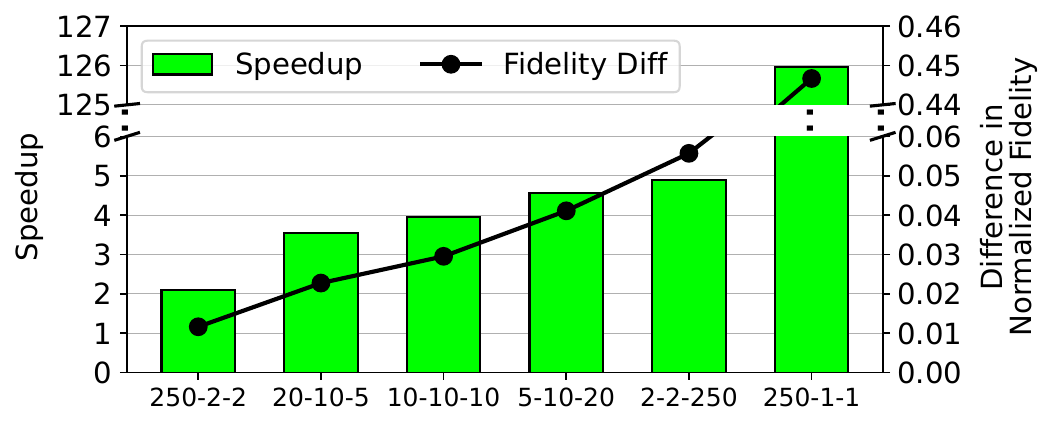}
    \caption{Speedups and difference in normalized fidelity for \texttt{QPE\_9} with 6 \qtree tree structures as compared to baseline. In extreme cases, fidelity significantly differs when only the \texttt{A0} outcomes are produced.}%
    \label{fig:var_struct}%
\end{figure}

Figure~\ref{fig:var_struct} shows these six structures' speedups and mean normalized fidelity differences. The results demonstrate a strong correlation between speedup potential and output accuracy, where even modest performance improvements lead to significant accuracy degradation. This is particularly evident in the extreme case where producing solely \texttt{$A_0$} outcomes results in substantial deviation from the baseline. \qtree's DCP approach maintains high accuracy while achieving performance benefits, which is crucial for ensuring reliable noisy simulation results.

\subsection{ Simulation Accuracy of VQAs with \qtree}

\begin{figure}[h!]
    \centering
    \includegraphics[width=0.8\columnwidth]{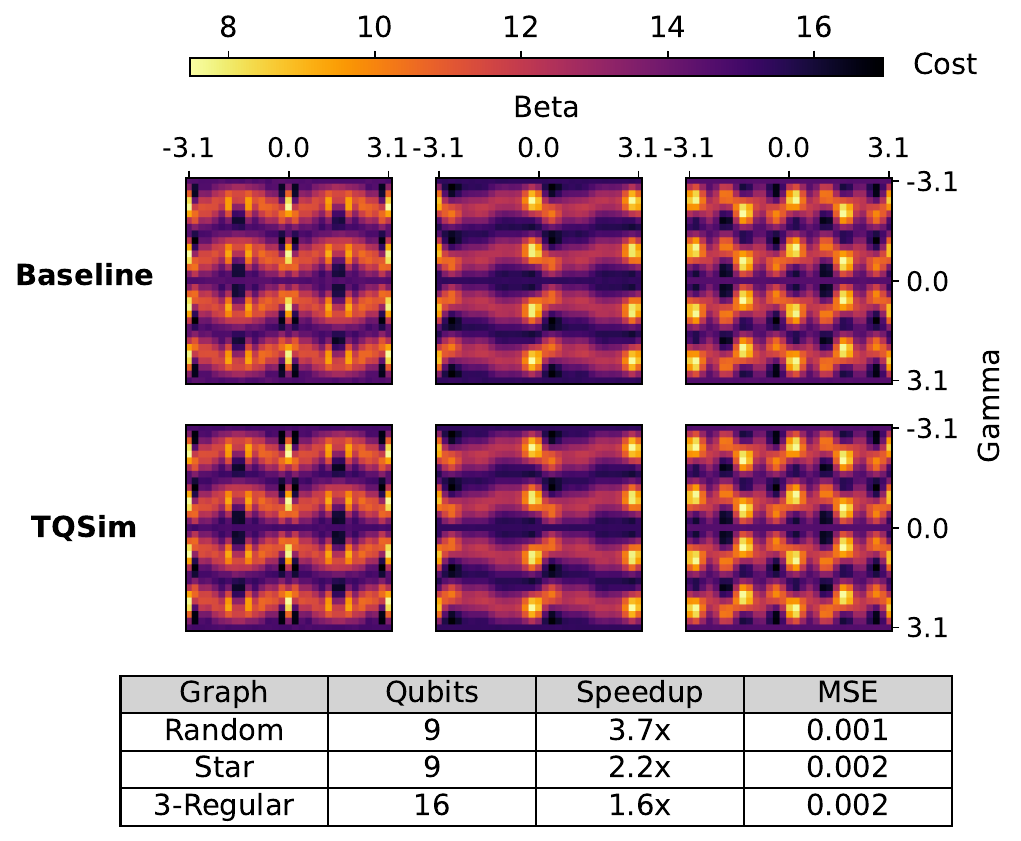}
    \caption{Cost function landscapes of the QAOA circuit designed to solve the max-cut problem for three input graphs.}
    \label{fig:qaoa_casestudy}
\end{figure}

\emph{Variational Quantum Algorithms} (VQAs) are pivotal near-term applications applicable to chemistry, optimization, and machine learning. These algorithms utilize parametric circuits and classical optimizers to refine circuit parameters for enhanced solution quality. However, designing VQAs compatible with existing noisy quantum computers remains a challenge. Currently, researchers heavily rely on noisy simulators for VQA design and evaluation. Unfortunately, each VQA iteration requires running the circuit numerous times, resulting in millions of simulations to tune parameters. %

For example, the cost landscapes shown in Figure~\ref{fig:qaoa_casestudy} are used to study the impact of noise on a 16-qubit QAOA circuit. To generate such landscapes, we need to run a total of 961 circuits by varying $\beta_0$ and $\gamma_0$ with depolarizing noise and an error probability of 0.1\%. The \emph{baseline simulator required 10.3 hours to run a grid search} on two circuit parameters, $\beta_0$ and $\gamma_0$. \emph{\qtree ran this task in about 6.4 hours, providing a 1.61$\times$ speedup}. The second row in Figure~\ref{fig:qaoa_casestudy} shows the output landscape produced by \qtree, which is almost identical to the baseline as the average mean squared error (MSE) between the two is 0.00161. Moreover, we observe similar gains without loss of accuracy on several other QAOA benchmarks~\cite{QAOAKit}.

\section{Related Work}
\label{Related-Work}

\noindent\textbf{1. Ideal Simulation Optimization:} HyQuas~\cite{HyQuas} partitions the quantum circuit into multiple subcircuits by its depth and uses the most suitable simulation technique for each subcircuit to enable speedup for ideal simulation. CutQC~\cite{cutqc} and Clifford circuit cutting~\cite{smith2023clifford} partition the quantum circuit into subcircuits along its width such that each subcircuit can be executed on a smaller quantum computer (or simulated more efficiently on a classical computer) and can be post-processed to generate complete output. Although both HyQuas and CutQC partition the circuit into subcircuits, their main goal is to optimize for the single-shot simulation, whereas \qtree focuses on multi-shot simulation optimizations. 

Furthermore, prior work on quantum circuit simulation uses gate fusion, data-level, and thread-level parallelism, and improves cache utilization to reduce the simulation time ~\cite{hpca_faster_shrodiner_style}. Tools like qHiPSTER~\cite{qHiPSTER}, QuEST~\cite{jones2019quest}, SV-Sim~\cite{li2021sv}, and DM-Sim~\cite{dmsim} support multi-node HPC simulation with lean, heterogeneous-ready infrastructures. Several works focus on reducing the memory requirement of quantum circuit simulation by using tensor networks, knowledge graphs, decision trees, and data compression~\cite{bsd,full_state_compression,noisy_variational_simulation,li2021sw_qsim,haner2016high,pang2020efficient}. Beyond general-purpose simulators, a class of quantum circuit simulators can simulate special quantum circuits efficiently~\cite{stablizer_simulator, extended_stablizer,MPS}. These works that use state vector simulation methods primarily focus on ideal quantum circuits that sample from a single probability distribution~\cite{cutqc,hpca_faster_shrodiner_style,HyQuas, qHiPSTER}. \qtree can be combined with these techniques further to improve the overall multi-shot quantum circuit simulation performance.

\begin{figure}
    \centering
    \includegraphics[width=0.8\linewidth]{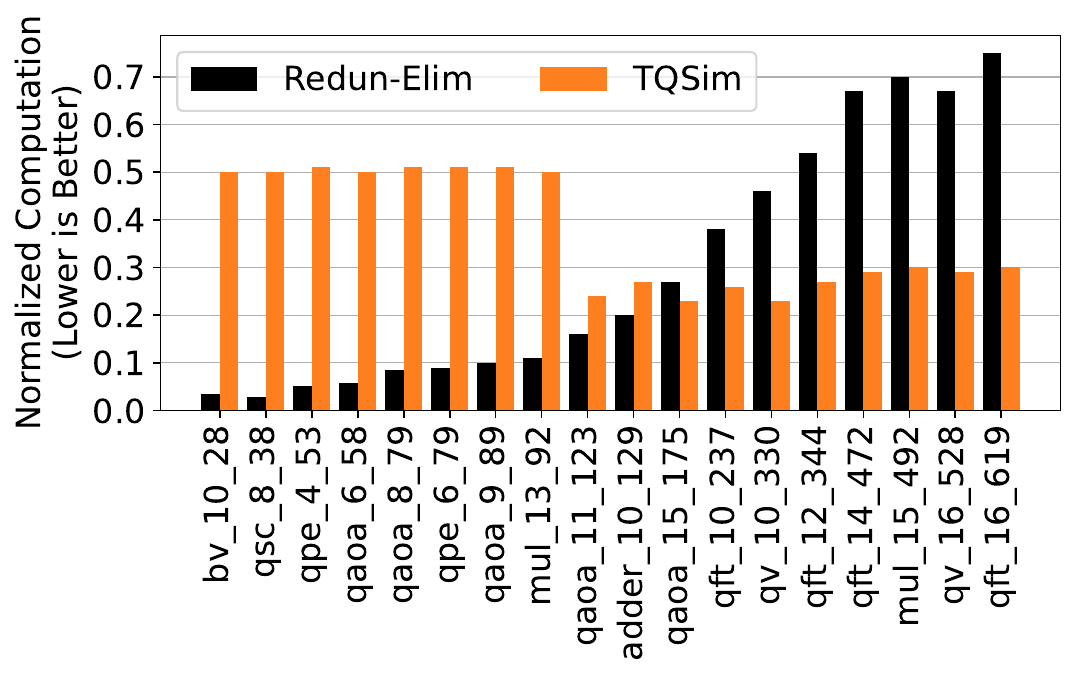}
    \caption{The normalized computation of the redundancy elimination (Redun-Elim) method as compared to \qtree. Noise is modeled using the depolarizing channel. The x label shows the benchmark name, width, and the number of gates.}
    \label{fig:dac-comp}
\end{figure}

\noindent\textbf{2. Inter-Shot Noisy Simulation Optimization:} Doi et al.~\cite{horii2023efficient} focus on scheduling and batch-processing to optimize GPU accelerations for multi-shot quantum circuit simulations. Li et al.~\cite{DAC-Redundancy} proposed an inter-shot redundancy elimination technique. Among all the noisy-version circuits, it searches for the \emph{identical} circuit portions and reuses the intermediate states for those redundancies. The experiment results show that this method can eliminate up to 90\% of computations for small circuits. However, the ratio of absolute redundancy drops significantly as the gate count increases. With the linear increase in the number of gates, the possible combination of error operators grows polynomially, and the chance of finding two exact sequences of errors becomes negligible. 

For example, using error rates from Google’s Sycamore machine, the redundancy ratio drops significantly as the gate count exceeds 200. Figure~\ref{fig:dac-comp} shows the experiment results of the redundancy elimination method applied to the benchmarks used in this paper. For circuits with less than 150 gates, the redundancy elimination method outperforms \qtree, but for circuits with greater than 150 gates, \qtree outperforms the redundancy elimination method.

\noindent\textbf{3. Application Specific Simulation Optimization:} Variational quantum algorithms (VQA) are viewed as one of the key quantum algorithms in the NISQ era. A series of optimizations have been proposed to refine their hybrid classical-quantum execution flow. These enhancements aim to optimize the execution process~\cite{wang2023enabling} and enable more effective simulations of such algorithms~\cite{lykov2023fast}. The insights from \qtree can be combined with those optimization methods to enhance these application-specific simulators further.

\section{Conclusion}
\label{Conclusion}
Quantum circuit simulation sees a significant slowdown when noise is incorporated into the simulation. To address this, we propose \qtree, a quantum circuit simulator designed to mitigate the slowdown induced by noise in simulations. \qtree leverages computational reuse by utilizing intermediate results across multiple shots, dynamically partitions subcircuits, and employs an efficient shot-allocation method. It is optimized to run on CPUs, GPUs, and multiple nodes, enhancing memory utilization on HPC clusters for significant speedups. Our experiments demonstrate up to $3.89\times$ faster performance than existing noisy quantum circuit simulators on a single node, with tight fidelity bounds.

\begin{acks}
This work was supported by the National Research Council (NRC) Canada grant AQC 003, AQC 213, and the Natural Sciences and Engineering Research Council of Canada (NSERC) [funding number RGPIN-2019-05059]. Swamit Tannu was supported by NSF Awards SHF:FET\#2212232 and QuSeC-TAQS:\#2326784. We would like to thank Tianyi Hao for providing valuable feedback. We also thank Rui Huang for reviewing an earlier draft of this paper and providing insightful comments, particularly on QAOA use cases. This research used the National Energy Research Scientific Computing Center (NERSC) resources, a U.S. Department of Energy Office of Science User Facility located at Lawrence Berkeley National Laboratory, operated under Contract No. DE-AC02-05CH11231. 
\end{acks}

\balance
\bibliographystyle{ACM-Reference-Format}
\bibliography{refs}

\appendix
\section{Artifact Appendix}

\subsection{Abstract}

\qtree is a noisy quantum circuit simulator. The main artifact is a C++ library that uses Qulacs as the simulation backend. Additionally, a Python front end streamlines experiment execution by processing user input, generating commands to run the C++ executable, collecting results, and plotting the output. There are two key metrics used in the paper to evaluate \qtree and they are: speedup (Figure~\ref{fig:speedups}) and accuracy (Figure~\ref{fig:norm-fidelity-diff}). 

\subsection{Artifact check-list (meta-information)}
{\small
\begin{itemize}
 \item {\bf Algorithm:} TQSim.
 \item {\bf Program:} C++ program using Qulacs as a backend with Python frontend for user interaction.
 \item {\bf Compilation:} GCC, Makefile, CMake.
 \item {\bf Model:} N/A.
 \item {\bf Data set:} 48 benchmark quantum circuits across 8 algorithm classes.
 \item {\bf Run-time environment:} Python interpreter and C++ runtime.
 \item {\bf Hardware:} Standard x86\_64 CPU, no specialized hardware required.
 \item {\bf Run-time state:} N/A.
 \item {\bf Execution:} C++ executable for Linux, compiled with GCC.
 \item {\bf Metrics:} Simulation speedup and fidelity accuracy.
 \item {\bf Output:} Reproduced Figure~\ref{fig:speedups} and Figure~\ref{fig:norm-fidelity-diff}.
 \item {\bf Experiments:} 48 noisy quantum circuit simulation jobs across varying qubit counts.
 \item {\bf How much disk space required (approximately)?:} <100 MB.
 \item {\bf How much time is needed to prepare workflow (approximately)?:} 10-15 minutes.
 \item {\bf How much time is needed to complete experiments (approximately)?:} 4–6 hours for a pre-selected subset of benchmarks with up to 13 qubits (depending on hardware). Benchmarks involving higher qubit counts require significantly more time.
 \item {\bf Publicly available?:} Yes, at \url{https://github.com/meng-ubc/TQSim}
 \item {\bf Code licenses (if publicly available)?:} MIT license
 \item {\bf Data licenses (if publicly available)?:} N/A
 \item {\bf Workflow automation framework used?:} N/A
 \item {\bf Archived (provide DOI)?:} \url{https://doi.org/10.5281/zenodo.15104095}
\end{itemize}
}

\subsection{Description}
TQSim is a tree-based reuse-focused noisy quantum circuit simulator that uses Qulacs as its backend. 

\subsubsection{How to access}
The artifact is available on GitHub at \url{https://github.com/meng-ubc/TQSim} or can be downloaded from Zenodo at \url{https://doi.org/10.5281/zenodo.15104095}. A README file is included with the code that provides detailed instructions for reproducing the paper's results.

\subsubsection{Hardware dependencies}
The artifact has been tested on standard x86\_64 hardware. No specialized hardware is required, though performance will improve with more CPU cores.

\subsubsection{Software dependencies}
\begin{itemize}
   \item C++ compiler (GCC recommended)
   \item Boost $\geq$ 1.71.0
   \item CMake $\geq$ 3.0
   \item Python $\geq$ 3.7
   \item Python packages: tqdm, matplotlib, numpy
\end{itemize}

\subsubsection{Data sets}
No external datasets are required. The artifact includes 48 benchmark quantum circuits across 8 algorithm classes, each with 6 circuits (see Table~\ref{table:benchmarks} for details).

\subsubsection{Models}
The artifact implements the tree-based reuse-focused simulation model described in the paper.

\subsection{Installation}
Clone and compile both the Qulacs dependency and TQSim:

\begin{verbatim}
# Install Qulacs
git clone https://github.com/qulacs/qulacs.git
cd qulacs
./script/build_gcc.sh
cd ..

# Install TQSim
git clone https://github.com/meng-ubc/TQSim.git
cd TQSim
git checkout AE
mkdir out
make
\end{verbatim}

Note that TQSim and Qulacs must be in the same directory as the makefile uses relative paths to locate the Qulacs library.

\subsection{Experiment workflow}
The evaluation workflow consists of two main steps:

\begin{enumerate}
   \item Run benchmarks to generate performance data:
   \begin{verbatim}
   cd ae_scripts
   python get_results.py
   \end{verbatim}
   
   \item Generate plots to visualize the results:
   \begin{verbatim}
   python plot.py
   \end{verbatim}
\end{enumerate}

The default configuration runs a subset of benchmarks (circuits with $\leq$ 13 qubits) to ensure reasonable execution time while still covering all benchmark classes.

\subsection{Evaluation and expected results}
Running the evaluation produces two key figures:
\begin{itemize}
   \item \texttt{Figure\_11\_speedup.png}: Reproduces Figure 11 (a)-(h) from the paper, showing the speedup achieved by TQSim compared to baseline simulators across the 8 benchmark classes.
   \item \texttt{Figure\_14\_fidelity.png}: Reproduces Figure 14 from the paper, demonstrating the fidelity comparison with the depolarizing noise model.
\end{itemize}

The results should show that TQSim achieves an average 2.51$\times$ speedup across all benchmark classes, with varying degrees depending on the circuit structure. The fidelity results should match those reported in the paper.

\subsection{Experiment customization}
Individual benchmarks can be run with specific configurations:

\begin{verbatim}
python get_results.py [class] [index]
\end{verbatim}

For example, to run the 15-qubit multiplier circuit:

\begin{verbatim}
python get_results.py mul 1
\end{verbatim}

This will update \texttt{results.json} with the new results, which can then be visualized using \texttt{plot.py}.

\subsection{Notes}
Running the complete set of 48 benchmarks requires significant computational resources and time, particularly for the larger circuits (20+ qubits). We recommend starting with the default configuration and selectively adding larger benchmarks as needed.

\subsection{Methodology}

Submission, reviewing and badging methodology:

\begin{itemize}
  \item \url{https://www.acm.org/publications/policies/artifact-review-and-badging-current}
  \item \url{https://cTuning.org/ae}
\end{itemize}

\end{document}